\documentclass[aps,12pt,superscriptaddress,nofootinbib,floatfix]{revtex4}
\usepackage{latexsym}
\usepackage{amsfonts}
\usepackage[latin1]{inputenc} 
\usepackage{subfigure} 
\usepackage{graphicx}
\usepackage{mathrsfs}
\usepackage{amssymb}
\usepackage{amsmath}
\usepackage{verbatim}
\usepackage{axodraw}

\def\mpt{{\slash\!\!\!\!\!\:P}_T}

\def\mptv{{\slash\!\!\!\!\!\:\vec{P}}_T}

\begin{document}

\begin{flushright}
{SHEP-08-13}\\
\today
\end{flushright}

\title{Phenomenology of the minimal $B-L$ extension\\[0.15cm]
of the Standard model: $Z'$ and neutrinos}

\author{Lorenzo Basso$^{1,2}$,
     Alexander Belyaev$^{1,2}$, 
     Stefano Moretti}
\affiliation{
School of Physics \& Astronomy, University of Southampton,\\
Highfield, Southampton SO17 1BJ, UK}
\affiliation{
Particle Physics Department, Rutherford Appleton Laboratory, \\Chilton,
Didcot, Oxon OX11 0QX, UK
}
\author{Claire H. Shepherd-Themistocleous}
\affiliation{
Particle Physics Department, Rutherford Appleton Laboratory, \\Chilton,
Didcot, Oxon OX11 0QX, UK
}
\begin{abstract}
\vspace*{1.5truecm} We present the Large Hadron Collider (LHC) discovery
potential in the $Z'$ and heavy neutrino sectors of a $U(1)_{B-L}$
enlarged Standard Model also encompassing three heavy Majorana
neutrinos. This model exhibits novel signatures at the LHC,
the most interesting arising from a $Z'$ decay chain involving
heavy neutrinos, eventually decaying into leptons and jets. In
particular, this signature allows one to measure the $Z'$ and heavy
neutrino masses involved.  In addition, over a large region of
parameter space, the heavy neutrinos are rather long-lived particles
producing distinctive displaced vertices that can be 
seen in the detectors.  Lastly, the simultaneous
measurement of both the heavy neutrino mass and decay length enables
an estimate of the absolute mass of the parent light neutrino.
\end{abstract}
\maketitle

\newpage
\section{Introduction}
{
The 
$B-L$ (baryon number minus lepton number) symmetry plays an important role in
various physics scenarios beyond the Standard Model (SM). Firstly, 
the gauged  $U(1)_{B-L}$ symmetry group is 
 contained in a  Grand Unified Theory (GUT) described by a $SO(10)$
group~\cite{Buchmuller:1991ce}.  Secondly, the scale of  the $B-L$
symmetry  breaking is related to the mass scale of the heavy right-handed 
Majorana neutrino mass terms providing the well-known see-saw
mechanism~\cite{see-saw} of light neutrino mass generation. Thirdly, the
$B-L$ symmetry and the scale of its breaking are tightly connected to
the baryogenesis mechanism through leptogenesis~\cite{Fukugita:1986hr}
via sphaleron interactions preserving $B-L$.

In the present paper we study the minimal $B-L$ low-energy extension
of the SM consisting of a further $U(1)_{B-L}$ gauge group, three
right-handed neutrinos and an additional Higgs boson generated through
the $U(1)_{B-L}$ symmetry breaking. It is important to note that in
this model the ${B-L}$ breaking can take place at the Electro-Weak
(EW) or TeV scale, i.e., a value far below that of any GUT scale.  
This $B-L$ scenario therefore has potentially interesting signatures at
hadron colliders, particularly the LHC. New particle states such as 
$Z'$, Higgses and neutrinos, all naturally have 
masses at the EW or TeV scale. 
The breaking of the $B-L$ symmetry at the EW or TeV scale can be
viewed as a remnant of a grand unified gauge symmetry, such as
$SO(10)$.
Furthermore, with respect to baryogenesis, since $B + L$ is violated
by sphaleron interactions, this implies that baryogenesis or
leptogenesis cannot occur above the scale of $B-L$ breaking. A
scenario with $B-L$ breaking at the EW (or TeV) scale therefore
implies EW (or TeV) scale baryogenesis~\cite{Trodden:1998ym}.

The particular subject of the present paper is the first detailed
study of the collider phenomenology of the gauge and fermionic sectors
of the minimal $B-L$ extension of the SM, where the additional $U(1)_{B-L}$
gauge group is indeed associated to the $B-L$ number
\cite{Buchmuller:1991ce,King:2004cx,B-L}.  The analysis of the scalar sector 
will appear in a future paper \cite{InProgress}. The new results on
$B-L$ phenomenology at the LHC include observable signals from a
$Z'$-boson as well as heavy neutrinos with a mass of up to several hundred
GeV\footnote{We deliberately assume that the $B-L$ symmetry breaking
scale generating $M_{Z'}$ is somewhat higher than the scale of the
heavier right-handed neutrino mass, thereby enabling $Z'$ to heavy
neutrino decays.}. A very interesting feature of such a $B-L$ model is
possibly relatively long lifetimes of the heavy neutrinos which can
directly be measured. In turn, such a measurement could be a key to
sheding light on the mass spectra of the light neutrinos.}

This work is organised as follows. Sect.~\ref{sect:model} reviews
the model under study and its implementation in the CalcHEP package
\cite{calchep}, together with an overview of its parameter
space. Sects.~\ref{sect:Hn} and \ref{sect:Zp} study the decay
properties -- specifically, width and Branching Ratios (BRs) -- of the
new spin-1 and spin-1/2 particles of the $B-L$ model. 
Sect.~\ref{sect:num_analys} discusses their experimental signatures
and production plus decay cross sections and also contains a numerical
analysis for two particular benchmark points in the $B-L$ parameter
space, followed by a study of the expected background. The 
conclusions are in Sec.~\ref{sect:conclusions}.

\section{The $\boldsymbol{B-L}$ model and its implementation into CalcHEP}\label{sect:model}
\subsection{The model}

The model under study is the so-called ``pure'' or ``minimal''
$B-L$ model (see \cite{B-L} for conventions and references) 
since it has vanishing mixing between the two $U(1)_{Y}$ 
and $U(1)_{B-L}$ groups.
In the rest of this paper we refer to this model simply as the ``$B-L$
model''.  {In this model the} classical gauge invariant Lagrangian,
obeying the $SU(3)_C\times SU(2)_L\times U(1)_Y\times U(1)_{B-L}$
gauge symmetry, can be decomposed as:
\begin{equation}\label{L}
\mathscr{L}=\mathscr{L}_{YM} + \mathscr{L}_s + \mathscr{L}_f + \mathscr{L}_Y \, .
\end{equation}
The non-Abelian field strengths in $\mathscr{L}_{YM}$ are the same as in the SM
whereas the Abelian
ones can be written as follows:
\begin{equation}\label{La}
\mathscr{L}^{\rm Abel}_{YM} = 
-\frac{1}{4}F^{\mu\nu}F_{\mu\nu}-\frac{1}{4}F^{\prime\mu\nu}F^\prime _{\mu\nu}\, ,
\end{equation}
where
\begin{eqnarray}\label{new-fs3}
F_{\mu\nu}		&=&	\partial _{\mu}B_{\nu} - \partial _{\nu}B_{\mu} \, , \\ \label{new-fs4}
F^\prime_{\mu\nu}	&=&	\partial _{\mu}B^\prime_{\nu} - \partial _{\nu}B^\prime_{\mu} \, .
\end{eqnarray}
In this field basis, the covariant derivative is:
\begin{equation}\label{cov_der}
D_{\mu}\equiv \partial _{\mu} + ig_S T^{\alpha}G_{\mu}^{\phantom{o}\alpha} 
+ igT^aW_{\mu}^{\phantom{o}a} +ig_1YB_{\mu} +ig_1'Y_{B-L}B'_{\mu}\, .
\end{equation}
The fermionic Lagrangian (where $k$ is the
generation index) is given by
\begin{eqnarray} \nonumber
\mathscr{L}_f &=& \sum _{k=1}^3 \Big( i\overline {q_{kL}} \gamma _{\mu}D^{\mu} q_{kL} + i\overline {u_{kR}}
			\gamma _{\mu}D^{\mu} u_{kR} +i\overline {d_{kR}} \gamma _{\mu}D^{\mu} d_{kR} +\\
			  && + i\overline {l_{kL}} \gamma _{\mu}D^{\mu} l_{kL} + i\overline {e_{kR}}
			\gamma _{\mu}D^{\mu} e_{kR} +i\overline {\nu _{kR}} \gamma _{\mu}D^{\mu} \nu
			_{kR} \Big)  \, ,
\end{eqnarray}
 {with the respective fermion charges given in Tab.~\ref{tab:ferm_content}.
  The  $B-L$ charge assignments of new fields
  as well as the introduction of new
  scalar Higgs ($\chi$) and fermionic  right-handed heavy neutrinos ($\nu_R$)
  fields are designed to eliminate the triangle $B-L$  gauge anomalies.
  (Tab.~\ref{tab:scal_content} shows the scalar content and charges of our 
   $B-L$ model.)
  Therefore, the $B-L$  gauge extension of the SM group
  broken at the EW scale does necessarily require
  at least one new scalar field and three new fermionic fields which are
  charged with respect to the $B-L$ group. }

The scalar Lagrangian is:
\begin{equation}\label{new-scalar_L}
\mathscr{L}_s=\left( D^{\mu} H\right) ^{\dagger} D_{\mu}H + 
\left( D^{\mu} \chi\right) ^{\dagger} D_{\mu}\chi - V(H,\chi ) \, ,
\end{equation}
{with the scalar potential given by}
\begin{equation}\label{new-potential}
V(H,\chi ) = m^2H^{\dagger}H +
 \mu ^2\mid\chi\mid ^2 +
  \lambda _1 (H^{\dagger}H)^2 +\lambda _2 \mid\chi\mid ^4 + \lambda _3 H^{\dagger}H\mid\chi\mid ^2  \, ,
\end{equation}
{where $H$ and $\chi$ are the complex scalar Higgs 
doublet and singlet fields, respectively.}

Finally, the Yukawa interactions are:
\begin{eqnarray}\nonumber
\mathscr{L}_Y &=& -y^d_{jk}\overline {q_{jL}} d_{kR}H 
                 -y^u_{jk}\overline {q_{jL}} u_{kR}\widetilde H 
		 -y^e_{jk}\overline {l_{jL}} e_{kR}H \\ \label{L_Yukawa}
	      & & -y^{\nu}_{jk}\overline {l_{jL}} \nu _{kR}\widetilde H 
	         -y^M_{jk}\overline {(\nu _R)^c_j} \nu _{kR}\chi +  {\rm 
h.c.}  \, ,
\end{eqnarray}
{where $\tilde H=i\sigma^2 H^*$ and  $i,j,k$ take the values $1$ to $3$},
where the last term is the Majorana contribution
 and the others the usual Dirac 
ones.
%
\begin{table}[htb]
\begin{center}
\begin{tabular}{|c|c|c|c|c|}\hline
$ \psi $ & $ SU(3)_C$ & $ SU(2)_L$ & $ Y $ & $ {B-L} $\\ \hline &&&&\\
$q_L$ & $3$ & $2$ & $\displaystyle\frac{1}{6}$ & $\displaystyle\frac{1}{3}$\\ &&&&\\
$u_R$ & $3$ & $1$ & $\displaystyle\frac{2}{3}$ & $\displaystyle\frac{1}{3}$\\ &&&&\\
$d_R$ & $3$ & $1$ & $\displaystyle -\frac{1}{3}$ & $\displaystyle\frac{1}{3}$\\ &&&&\\
$l_L$ & $1$ & $2$ & $\displaystyle -\frac{1}{2}$ & $-1$\\ &&&&\\
$e_R$ & $1$ & $1$ & $-1$ & $-1$\\ &&&&\\
$\nu _R$ & $1$ & $1$ & $0$ & $-1$\\ \hline
\end{tabular}
\end{center}\caption{Fermion content and charges for the $B-L$ 
model\label{tab:ferm_content}.}
\end{table}
%
\begin{table}[htb]\begin{center}
\begin{tabular}{|c|c|c|c|c|}\hline
$ \psi $ & $ SU(3)_C$ & $ SU(2)_L$ & $ Y $ & $ {B-L} $\\ \hline &&&&\\
$H$ & $1$ & $2$ & $\displaystyle \frac{1}{2}$ & $0$\\ &&&&\\
$\chi $ & $1$ & $1$ & $0$ & $2$\\ \hline 
\end{tabular}
\end{center}\caption{Scalar content and charges for the $B-L$ 
model\label{tab:scal_content}.}
\end{table}

%
%
\subsection{Model implementation into CalcHEP}

{We make use of the CalcHEP package~\cite{calchep} 
 to study the collider phenomenology of the $B-L$ model.}
For the derivation of the Feynman rules (see the Appendix for those pertaining
to the heavy neutrino interactions)
and for the straightforward implementation of the model in the
CalcHEP package, {
we have used the LanHEP module \cite{lanhep}. 
The availability of the model implementation into CalcHEP in both the
unitary and t'Hooft-Feynman gauges allowed us to perform powerful
cross-checks to test the consistency of the model itself}.

{The implementation of the gauge sector is quite
straightforward. Since there is no mixing between the (SM) $Z$ and
$Z_{B-L}$ bosons (hereafter, we will refer to the $Z_{B-L}$ boson as a
$Z'$) one just needs to define a new heavy neutral gauge boson
together with the covariant derivative given by eq. (\ref{cov_der})
and the charge assignments in
Tabs. \ref{tab:ferm_content}--\ref{tab:scal_content}. For the scalar
sector, we need to implement the mixing between mass and gauge
eigenstates of the two Higgs bosons.

The implementation of the neutrino sector is somewhat more
complicated.  Majorana-like Yukawa terms are present in
eq. (\ref{L_Yukawa}) for the right-handed neutrinos, therefore one
must implement this sector such that the gauge invariance of the model
is explicity preserved.  This can be done as follows.  As a first step
we rewrite Dirac neutrino fields in terms of Majorana ones using the
following general substitution:
\begin{equation}\label{D-M}
\nu^D = \frac{1-\gamma _5}{2}\nu_L +  \frac{1+\gamma _5}{2}\nu_R\, ,
\end{equation}
where $\nu^D$ is a Dirac field and $\nu_{L(R)}$ are its left (right) Majorana components. 
If we perform the substitution of eq. (\ref{D-M}) 
in the neutrino sector of
the SM, we will have an equivalent theory formulated in 
terms of Majorana neutrinos consistent with all
experimental constraints.}

{The second step is to diagonalise the neutrino mass matrix from eq. (\ref{L_Yukawa}):}
\begin{equation}\label{nu_mass_matrix} 
{\mathscr{M}} = 
\left( \begin{array}{cc} 0 & m_D \\ 
                  m_D &  M 
 \end{array} \right)\, , 
\end{equation} 
where
\begin{equation} 
m_D = \frac{y^{\nu}}{\sqrt{2}} \, v \, , \qquad M = \sqrt{2} \, y^{M} \, x \, ,
\end{equation}
{where $x$ is the Vacuum Expectation Value (VEV) of the $\chi$ field.
This matrix can be  diagonalised }
by a rotation about an angle $\alpha _\nu$, such that:
\begin{equation}\label{nu_mix_angle} 
\tan{2 \alpha_\nu} = -\frac{2m_D}{M}\, .
\end{equation}

{For simplicity we neglect the inter-generational mixing
so that neutrinos of each generation can be
diagonalised independently.
We also require that the neutrinos be mass 
degenerate. 
Thus, $\nu_{L,R}$ can be written as the following linear combination
 of Majorana mass eigenstates $\nu_{l,h}$ :}
\begin{equation}\label{nu_mixing} 
\left( \begin{array}{c} \nu_L\\ \nu_R \end{array} \right) = 
\left( \begin{array}{cc} 
\cos{\alpha _\nu} & -\sin{\alpha_\nu} \\ 
\sin{\alpha _\nu} &\cos{\alpha _\nu} 
\end{array} \right) \times \left( \begin{array}{c} \nu_l\\ \nu_h \end{array} \right)\, . 
\end{equation} 

The last subtle point is the way the Lagrangian has to be written, 
in particular the Majorana-like Yukawa
terms for the right-handed neutrinos (the last term in eq. 
(\ref{L_Yukawa})).
 In order 
{to explicitly preserve}
gauge invariance, this term has to be written, in two-component notation, as:
\begin{equation}  - y^M \nu ^c \frac{1+\gamma _5}{2} \nu \chi + \rm{h.c.}\, ,
\end{equation}
where $\nu$ is the Dirac field of eq. (\ref{D-M}), whose Majorana 
components $\nu_{L,R}$  {mix as in 
eq. (\ref{nu_mixing}).}

\subsection{Parameter space}
{
In this section we define the independent parameters of the $B-L$
model and their valid 
range\footnote{Since the scalar sector is not 
within the scope of the present study we do not
discuss the corresponding parameters, as one can
choose settings in parameter space such 
that the scalars are entirely decoupled
from the remaining particles. Here, we achieve this by requiring
$\lambda _1 = 3$, $\lambda _2 = 0.08$ and $\lambda _3 = 0.01$,
so that the scalars masses are
 $m_{h_1} \approx 600$ GeV and $m_{h_2}
\approx 1.5$~TeV, corresponding
{to our default benchmark parameters in the $Z'$ sector,
$g'_1=0.2$,
$M_{Z'}=1.5$~TeV.} The only exceptions to this will be made in
Sects.~IIIA and \ref{sect:Zp}, where we will temporarily
adopt other settings, to describe
the complete decay pattern of the $Z'$ and {heavy neutrinos}.}.
{
The set of parameters relevant to our study is the following one.}
\begin{itemize}
\item 
{
$g'_1$, the new $U(1)_{B-L}$ gauge  coupling.
Here, the absence of a Landau pole up to the GUT scale $Q_{GUT}=10^{16}$ GeV
implies  $g'_1 < 0.5$ from a Renormalisation Group Equation
(RGE) analysis \cite{CPW_AC,B-L}.
}
\item $M_{Z'}$, the new gauge boson mass.
An indirect constraint on $M_{Z'}$ comes 
from analyses at the Large Electron-Positron (LEP) collider
of Fermi effective four-fermions interactions
\cite{Carena}:
\begin{equation}\label{LEP_bound}
\frac{M_{Z'}}{g'_1} \geq 6\; \rm{TeV}\, .
\end{equation}
As we demonstrate below, 
this constraint will provide an upper bound to the $Z'$ production cross sections at the
LHC.
\item$M_{\nu _h}$, the heavy neutrino masses. We take them to be 
degenerate and relatively
       light, varying in the range $50$ GeV $< M_{\nu _h} < 500$ GeV.
\item $m_{\nu _l}$, the SM (or light) neutrino masses.  We use the
cosmological upper bound $\sum_l m_{\nu _l}<1$ eV. As we will see in
Sect. \ref{sect:Hn}, detectable displaced vertices may occur for
$m_{\nu _l} \lesssim 10^{-2}$ eV.
\end{itemize}

{
For this analysis of the collider phenomenology of our $B-L$ model 
we have chosen $M'_Z =
1.5$ TeV and $g'_1 = 0.2$ as a representative
point satisfying present experimental constraints
as well as two values for heavy neutrino masses: $M_{\nu _h}=200$ and 500 GeV.
Finally, we have fixed the light
neutrino mass to be $m_{\nu_l}=10^{-2}$ eV. For illustrative purposes
we take all neutrino masses, both light and heavy, to be degenerate.
}

\section{Phenomenology of the $B-L$ model}
\subsection{Heavy neutrino properties}\label{sect:Hn}
\subsubsection{Heavy neutrino decays}\label{sect:Hn-decay}
{
After the diagonalisation of the neutrino mass matrix realising the see-saw
mechanism,  we obtain three very light neutrinos ($\nu_l$), which are the SM-like
neutrinos, and three heavy neutrinos ($\nu_h$).
The latter have an extremely small mixing
with the  $\nu_l$'s thereby providing very small but non-vanishing 
couplings to  gauge and Higgs bosons (see the Appendix 
for  the Feynman rules involving heavy neutrino interactions)
which in turn enable the following $\nu_h$ decays:
 $\nu_h \rightarrow l^\pm W^\mp$,
 $\nu_h \rightarrow \nu_l  Z$,
 $\nu_h \rightarrow \nu_l h_1$,
 $\nu_h \rightarrow \nu_l h_2$
 as well as $\nu_h \rightarrow \nu_l Z'$
when these decay channels are kinematically allowed.}
\begin{figure}[ht]
\begin{center}
\includegraphics[angle=0,height=0.6\textwidth,width=0.80\textwidth]{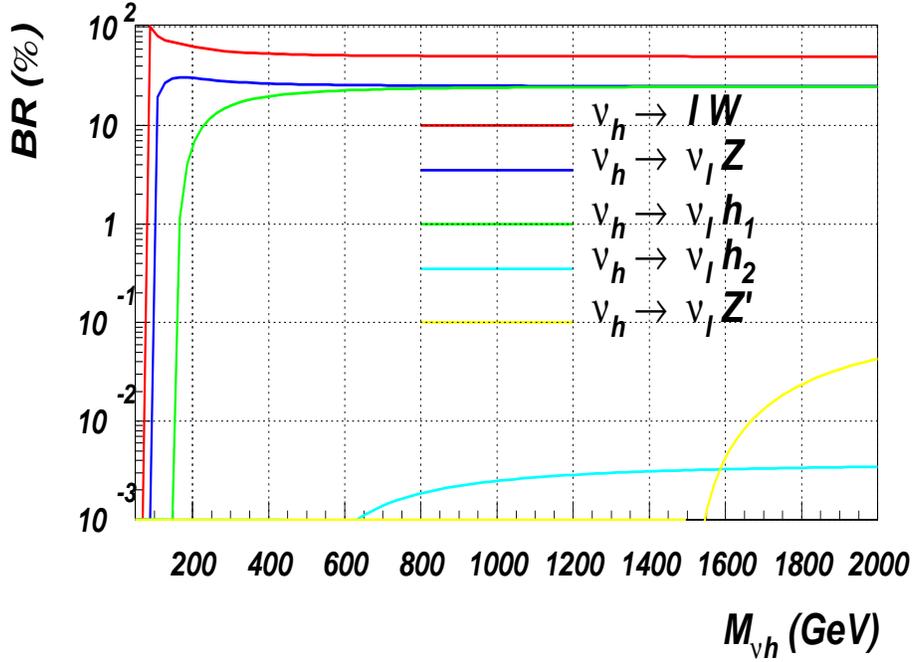}
\caption{Heavy neutrino branching ratios 
versus its mass for the fixed $M_{Z'}=1.5$~TeV,  
$m_{h_1}=150$ GeV and
$m_{h_2}=450$ GeV 
corresponding to 
$\lambda_1 = 0.19, \lambda_2=0.017, \lambda_3=0.01$
and $g'_1=0.2$.
Here, $W$ means the sum over $W^+$ and $W^-$.
\label{Hn_BR}
}
\end{center}
\end{figure}
{
 Fig.~\ref{Hn_BR} presents the corresponding BRs versus 
the heavy neutrino mass for the values of the other relevant $B_L$ parameters
given in the caption.
One can see that the $BR\left( \nu_h \rightarrow l^\mp W^\pm \right)$
is dominant and reaches the $2/3$ level in the  $M_{\nu_h} \gg M_W, M_Z$
limit, while 
$BR\left( \nu_h\rightarrow \nu_l Z \right)$
and
$BR\left( \nu_h\rightarrow \nu_l h_1\right)$
both reach the $1/6$ level in this regime.
In contrast, the $\nu_h \rightarrow \nu_l h_2$
as well as $\nu_h \rightarrow \nu_l Z'$
decay channels are well below the percent level
and are negligible for our study.}
In this paper we will eventually
 assume  that the heavy neutrino masses are smaller
than both Higgs boson masses. Under this assumption
$\nu_h\rightarrow \nu_l h_i$ ($i=1,2$) is not kimematically possible and 
$BR\left( \nu_h\rightarrow \nu_l Z\right)$
reaches the $1/3$ level in the  $m_{\nu_h} \gg M_W, M_Z$ limit.

\subsubsection{Lifetime of the heavy neutrinos\label{sect:Hn-life-time}}
{The heavy neutrino couplings to the weak gauge bosons are
proportional to the ratio of light and heavy neutrino masses (see the
Appendix), which is extremely small.  Therefore the decay width of
the heavy neutrino is correspondingly small and its lifetime large. }
The heavy neutrino can therefore be a long lived particle and, { over a
large portion of parameter space, its lifetime can be comparable to or
exceed that of the $b$-quark}.  (In fact,
for $m_{\nu_l}=10^{-2}$ eV and $M_{\nu_h}=200$ GeV they are equal.)
\begin{figure}[!ht]
\begin{center}
\includegraphics[angle=0,width=0.8\textwidth]{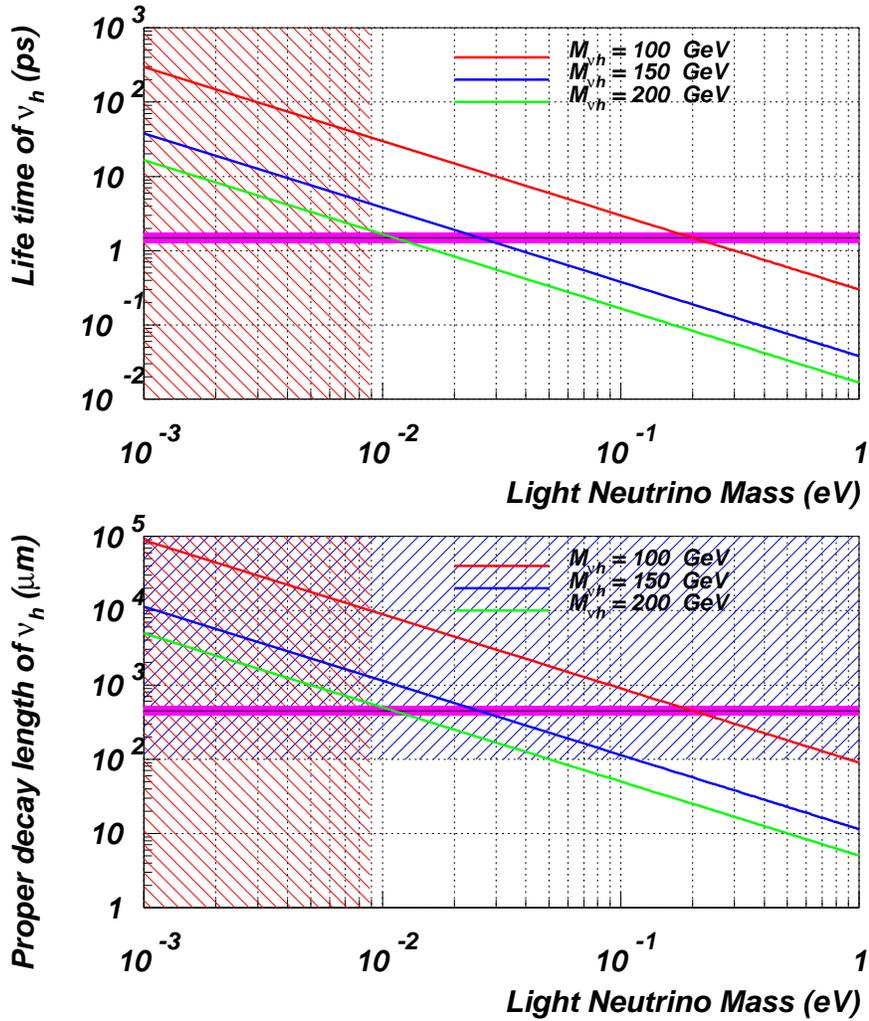}
\caption{
{Heavy neutrino life-time (top) and 
proper decay length (or mean path) $c\tau_0$ (bottom) as a function of the light neutrino
mass. The purple band presents the proper decay length of the $b$-quark
while  the blue 
band indicates the range of a typical
micro-vertex detector. The red band shows the region excluded by neutrino oscillation direct
measurements.}
}
\label{life_time}
\end{center}
\end{figure}
{In Fig. \ref{life_time} we present the heavy neutrino lifetime
(top) in pico-seconds and the {proper decay length} (or mean path)
(bottom) in micro-meters as a function of the light neutrino mass.
The proper decay length is defined as $c\tau_0$, where $\tau_0$ is
the lifetime of the heavy neutrino.  
{The purple band presents the
proper decay length of the $b$-quark while the blue band indicates the
range of a typical micro-vertex detector.}} The red band shows the
region of light neutrino masses excluded by direct measurements of
neutrino oscillations \cite{neutrino_mixing_angles/masses}, by taking
the lighter neutrino {to be massless} (so that the other neutrinos
cannot populate this region).  {One should also note that the
lifetime and the proper decay length of the heavy neutrinos in the
laboratory frame will actually be equal to those given in
Fig. \ref{life_time} times the Lorentz factor equal to
$p_{\nu_h}/M_{\nu_h}$ defined by the ratio $M_{Z'}/M_{\nu_h}$ which
can be as large as about a factor of 10.}  We can then see that there
exists a region where the heavy neutrino lifetime is of the same order
as {that of the $b$-quark (shown} as a purple band).  The mean path
and the respective lifetime of heavy neutrinos can therefore be
measured from a displaced vertex inside the detector.  
%
%
The heavy neutrino can however be distinguished from a $b$-hadron through the
observation of vertices consisting of only two isolated leptons. (A
SM $B$-meson decay while possible would have a very small BR, $\sim 10^{-8}$
at the most.)

An experimentally resolvable non-zero lifetime along with a mass
determination for the heavy neutrino also enables a determination of
the light neutrino mass. The lifetime measurement allows the small
heavy-light neutrino mixing to be determined and as one can see from 
eqs. (\ref{nu_mass_matrix})--(\ref{nu_mixing}) this, along with 
the heavy neutrino mass, gives the light neutrino mass. 
Considering only one generation for simplicity, this is expanded upon below.



Mass eigenstates are related to gauge eigenstates by
eq. (\ref{nu_mixing}), hence the eigenvalues are given by solving the
equation:
\begin{displaymath}
\left( \begin{array}{cc} m_{\nu _l} & 0 \\ 0 & M_{\nu_h} \end{array} \right) = \left( \begin{array}{cc} c_\nu 	& s_\nu \\ -s_\nu &c_\nu \end{array} \right)     \left( \begin{array}{cc} 0 & m_D \\ m_D^T & M \end{array} \right)     \left( \begin{array}{cc} c_\nu 	& -s_\nu \\ s_\nu &c_\nu \end{array} \right)\, ,
\end{displaymath}
which yields
\begin{eqnarray}\label{nul_mass}
m_{\nu _l} &=& \sin\, 2\alpha _\nu\, m_D + \sin^2 \alpha _\nu\, M\, , \\ \label{nuh_mass}
M_{\nu_h} &=& -\sin\, 2\alpha _\nu\, m_D + \cos^2 \alpha _\nu\, M\, ,
\end{eqnarray}
with $\alpha _\nu$ given by eq.~(\ref{nu_mix_angle}).
We have then three parameters ($m_D$, $M$ and $\alpha _\nu$) and a constraint
{(given by eq.~(\ref{nu_mix_angle})),
that can be used to eliminate one parameter from the above equations.

The Feynman rules given in the Appendix demonstrate that heavy
neutrino interactions are determined by the mixing angle $\alpha_\nu$
only,} as is the total width (and therefore the mean decay
length). Hence, it is convenient to keep $M_{\nu_h}$ and $\alpha_\nu$
{as independent model parameters} eliminating $m_D$ from
eq. (\ref{nu_mix_angle}),
\begin{equation}\label{elimin_1}
m_D = m_D(\alpha _\nu , M_{\nu_h})\, .
\end{equation}
By measuring the heavy neutrino mass we can also invert eq. (\ref{nuh_mass})
\begin{equation}\label{elimin_2}
M = M(\alpha _\nu, M_{\nu_h})\, ,
\end{equation}
to finally get a fully known expression for the SM light neutrino mass
as a function of our input parameters
$M_{\nu_h}$ and $\alpha _\nu$,
 that we can measure independently, by inserting eqs.
(\ref{elimin_1})--(\ref{elimin_2}) into eq. (\ref{nul_mass}),
\begin{equation}\label{SMnu_mass}
m_{\nu _l} (m_D,M) = m_{\nu_l}( \alpha _\nu, M_{\nu_h})\, .
\end{equation}
This {simple picture shows that  within the $B-L$ model
we have an indirect way of accessing the SM light neutrino mass by measuring 
the mass of the heavy neutrino and the kinematic 
features of its displaced  vertex. If the whole structure of
mixing is taken into account, including inter-generational
mixing in the heavy neutrino sector, the task of determining the
light neutrino mass this way would become more complicated
but the qualitative features and the overall strategy would
remain  the same thereby
providing one with a unique link between  very large and very small
mass objects.}

\subsection{$Z'$ decay properties}\label{sect:Zp} 
{
As discussed earlier, the extra $U(1)_{B-L}$  gauge
group provides an additional neutral gauge boson, $Z'$,
with no mixing with the SM $Z$-boson.
{Therefore our $Z'$ boson decays only to fermions
at tree-level and its width}  is given by 
the following expression:}
\begin{equation}\label{Z'_width}
\Gamma (Z'\rightarrow f\overline{f})=
\frac{M_{Z'}}{12\pi}C_f (v^f)^2 
\left[ 1 +2\frac{m_f^2}{M_{Z'}^2}\right]\sqrt{1-\frac{4m_f^2}{M_{Z'}^2}}\, ,
\end{equation}
where $m_f$ is the mass and $C_f$ the number of colours of the fermion type
 $f$ and $v^f = (B-L) \times g'_1$ is the vector coupling (see 
 Tab.~\ref{tab:ferm_content}).

{
In Figs. \ref{Zp_width_vs_MZp}  and  \ref{Zp_width_vs_g1p}
we present the total  decay width
of the $Z'$ as 
a function of $M_{Z'}$  and  $g'_1$, respectively (with the
other parameters held fixed to three
different values),
assuming that the partial decay width into heavy neutrinos vanishes.
Also, Fig. \ref{Zp_width_vs_Mhn_rel}  presents the relative variation of the total width as
a function of the $\nu_h$ mass for three different
values of  $M_{Z'}$ and with $g'_1 = 0.5$.
\begin{figure}[htb]
  \begin{center}
  \subfigure[]{ 
  \label{Zp_width_vs_MZp}
  \includegraphics[width=0.5\textwidth]{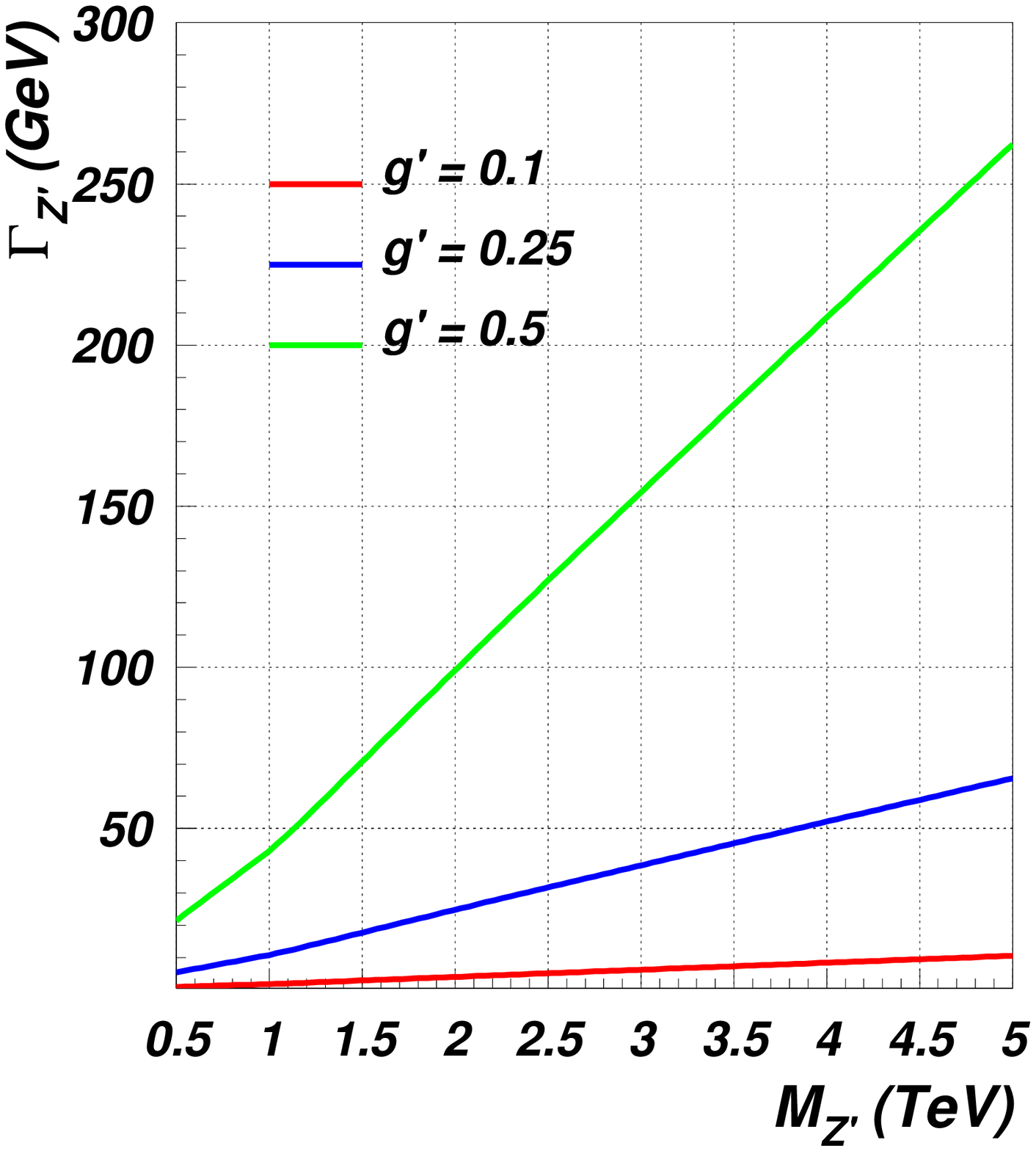}}%
  \subfigure[]{
  \label{Zp_width_vs_g1p}
  \includegraphics[width=0.5\textwidth]{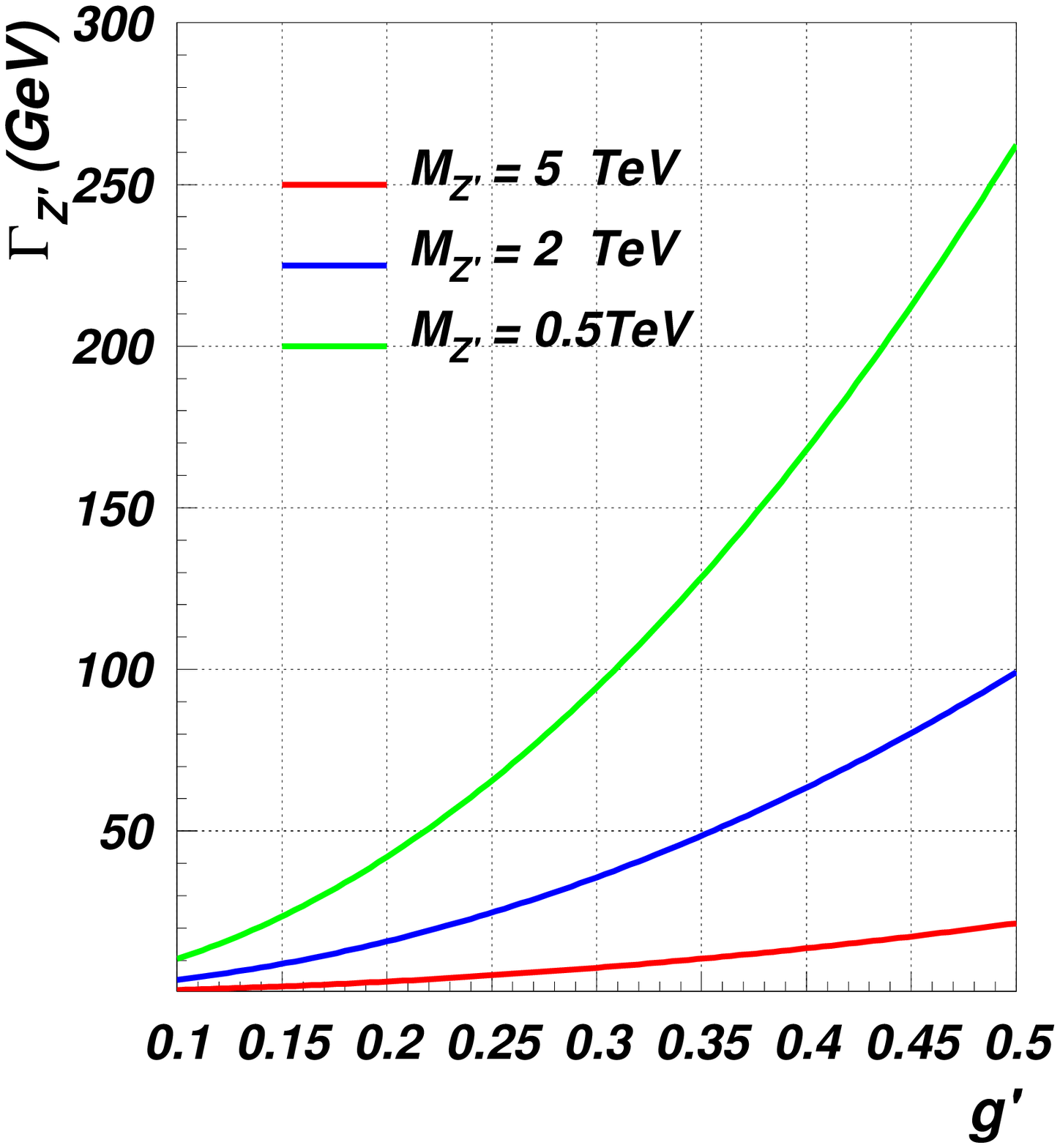}}
  \\
  \subfigure[]{ 
  \label{Zp_width_vs_Mhn_rel}
  \includegraphics[width=0.5\textwidth]{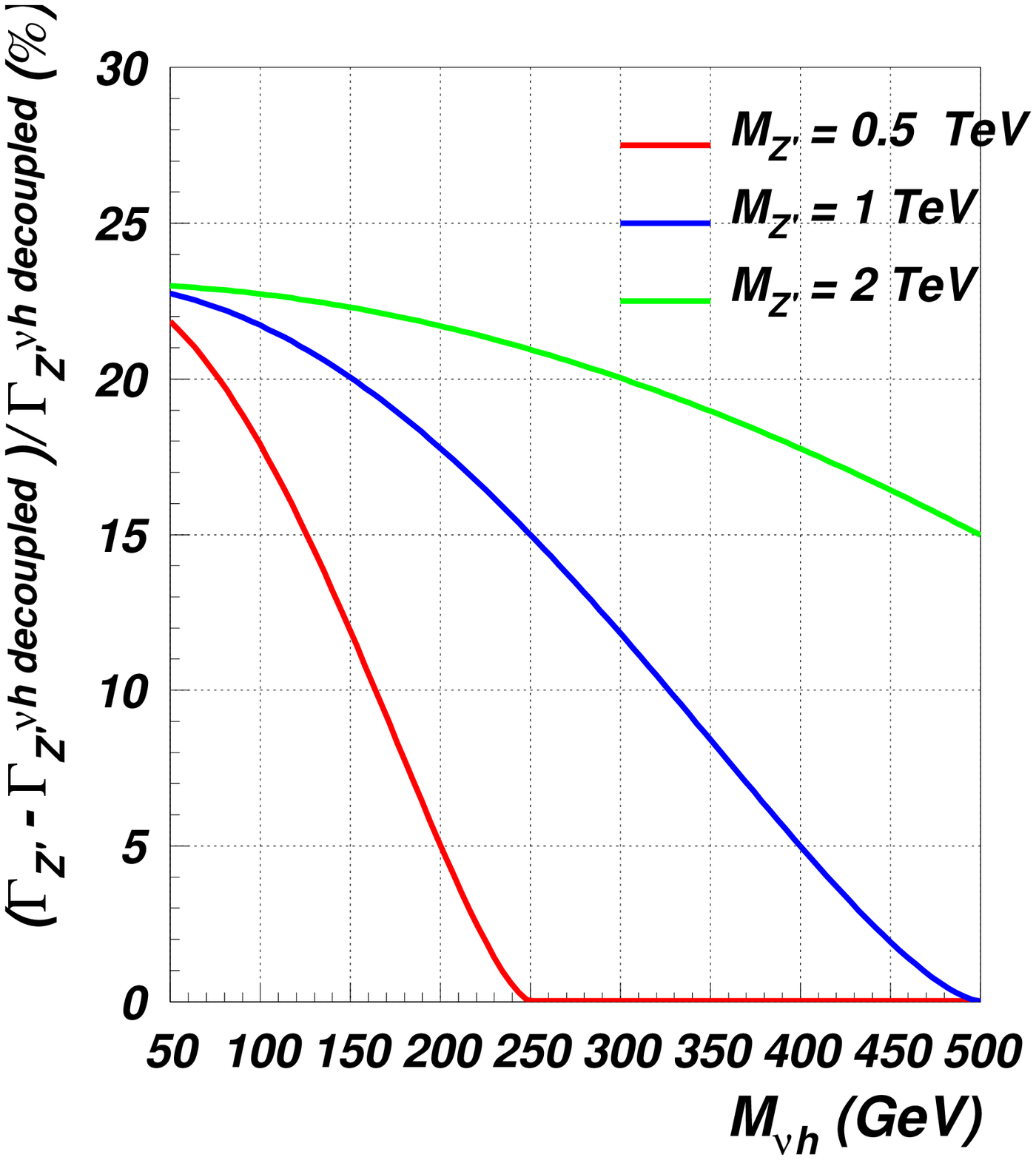}}%
  \end{center}
  \caption{$Z'$ total width as a function of: \ref{Zp_width_vs_MZp} $M_{Z'}$ (for fixed values  of $g'_1$), \ref{Zp_width_vs_g1p}
 $g'_1$ (for fixed values of $M_{Z'}$) and \ref{Zp_width_vs_Mhn_rel} $M_{\nu_h}$ (for fixed values of $M_{Z'}$ and $g'_1 = 0.5$).}
  \label{Zp_width}
\end{figure}

From the first two plots we see that the total width of a $Z'$ gauge
 boson varies from a few to hundreds of GeV over a mass range of $0.5$
 TeV $<M_{Z'}<5$ TeV, depending on the value of $g'_1$, while from the
 third plot one can gather the importance of taking into consideration the
 heavy neutrinos, since their relative contribution to the total width
 can be as large as $25\%$ (whenever this channel is open).
%
One should also note that possible $Z'$ decays into one light and one
heavy neutrino are highly suppressed by the corresponding
(heavy-light) neutrino mixing and thus they can safely be neglected.

The possibility of decays of the $Z'$ gauge boson into pairs of heavy
neutrinos { is one of the most significant results of this work}
since, in addition to the clean SM-like di-lepton signature, it
provides multi-lepton signatures where backgrounds can strongly be 
supressed. } In order to address this quantitatively, we first
determine the relevant BRs.  Clearly, these depend strongly
on the heavy neutrino mass and Fig. \ref{Zp_BR} shows how they change
with fixed (although arbitrary) values of $M_{\nu_h}$, for the
following three cases: a heavy neutrino (i) much lighter than, (ii) lighter
than and (iii) comparable in mass to the $Z'$, in the range $0.5$ TeV
$<M_{Z'}<5$ TeV, before summing over generations.

\begin{figure}[!ht]
  \centering
  \includegraphics[angle=0,width=1\textwidth]{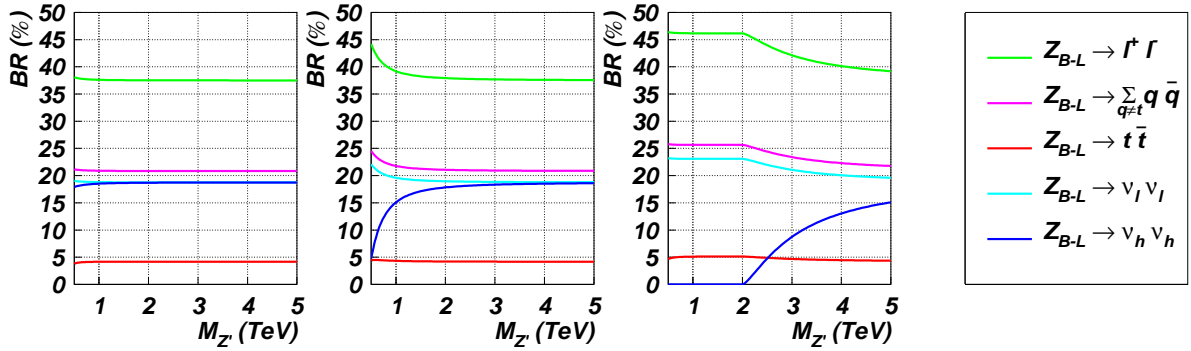}
  \caption{$Z'$ BRs as a function of $M_{Z'}$ for several heavy neutrino masses: $M_{\nu_h} =
50$, $200$ and $1000$ GeV, from left to right, respectively.
A summation over all lepton/neutrino flavours is implied throughout
whereas in the case of quarks we distinquish between light flavours
($q=d,u,s,c,b$) and the top quark.}
  \label{Zp_BR}
\end{figure}

A feature of the current $B-L$ model illustrated in the
previous figures is that the $Z'$ predominantly couples to
leptons. In fact, after summing over the generations, $k=1...3$, we
roughly get for leptons and quarks:
\begin{displaymath}
\sum _k BR\left( Z' \rightarrow l_k \overline{l_k} + \nu_k \overline{\nu_k} \right) \sim \frac{3}{4}\, ,\qquad
\sum _k BR\left( Z' \rightarrow q_k \overline{q_k} \right) \sim \frac{1}{4}\, .
\end{displaymath}
Not surprisingly then,  for a
relatively light (with respect to the $Z'$ gauge boson) heavy neutrino, the $Z'$ BR into
pairs of such particles is relatively high: $\sim 18\% $ (at most, again, 
after summing over the
generations).

Combining this last result together with those of Sect. \ref{sect:Zp}, we can
discuss an interesting feature of the $B-L$ model, namely,  the multi-lepton
signatures (meaning two or more leptons being involved). 
A single heavy
neutrino decay will produce a signature of 0, $1$ or $2$ charged leptons, 
depending on whether the heavy neutrino {decays} via a charged or neutral
current and on the subsequent decays of the SM $W^\pm$ and $Z$ gauge bosons. We
can have both
{chains}
\begin{equation}\label{Hn-W}
\nu_h \rightarrow l^\pm\, W^\mp \rightarrow l^\pm +\left\{ \begin{array}{c} l^\mp\,\nu _l\\ 
{\rm hadrons} \end{array}\right.
\end{equation}
{and}
\begin{equation}\label{Hn-Z}
\nu_h \rightarrow \nu _l\, Z \rightarrow \nu _l +\left\{ \begin{array}{c} l^+l^-\\ \nu_l\,\nu_l 
/{\rm hadrons}   
\end{array} \right.\, .
\end{equation}
The pattern in (\ref{Hn-W}) provides $1$ or $2$ charged leptons whilst
that in (\ref{Hn-Z}) zero or 2, so that multi-lepton signatures may
arise when the $Z'$ gauge boson decays into a pair of heavy neutrinos,
producing up to four charged leptons in the final state. Fig.
\ref{3-4lept} shows the BRs of a $Z'$ decaying into $2$
 (top-left) and $3$ or
$4$ (top-right) leptons (plus possibly missing transverse momentum and/or jets,
as appropriate)
as a function of $M_{\nu_h}$, 
where a lepton
can be either an electron or a muon and these contributions are
summed.  While the former are clearly dominant the latter are not at
all negligible.
\begin{figure}[!t]
\begin{center}
\includegraphics[angle=0,width=\textwidth]{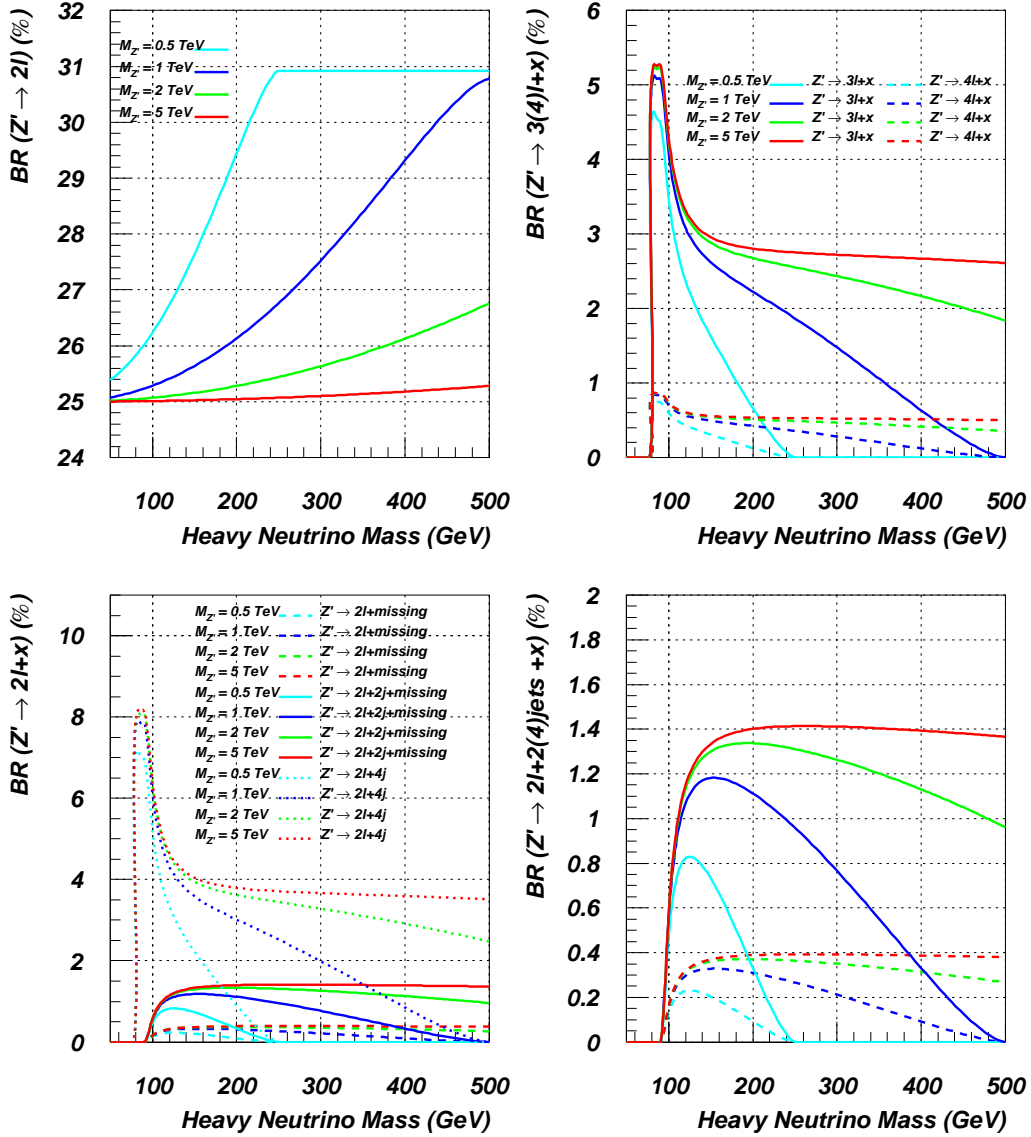}
\vspace*{-2truecm}
\caption{$Z'$ BRs, as a function of $M_{\nu_h}$, into: $2$ leptons (both
$e$ and $\mu$, top-left); $3$ and $4$ leptons + $X$
(both $e$ and $\mu$, top-right);
$2$ leptons + $X$ jets (both $e$ and $\mu$, bottom-left); zoom of
the previous plot with same legend (bottom-right).}
\label{3-4lept}
\end{center}
\end{figure}
For $M_{W^\pm}<M_{\nu_h}<M_Z$, the $\nu_h \rightarrow l^\mp\, W^\pm$
decay is the only one kinematically possible whereas for
$M_{\nu_h}<M_{W^\pm}$ the heavy neutrino can decay only via an off
shell W and is therefore very long lived.  For a very massive $Z'$
($2$ TeV $< M_{Z'}< 5$ TeV) the multi-leptonic BRs are roughly $2.5\%
$ in the case of $Z' \rightarrow 3l$ and $0.5\% $ in the case of $Z'
\rightarrow 4l$, for a wide range of heavy neutrino masses.

Finally, from (\ref{Hn-W}) and (\ref{Hn-Z}) one can see that 
di-lepton decays are
possible, whereas $Z'$ decays give rise to 2 leptons {plus} a
large amount of missing transverse momentum and/or highly energetic
jets: see Fig.~\ref{3-4lept} (bottom-left). Particularly interesting
is the decay into $2$ leptons and $4$ jets, since here there is no
missing transverse momentum at all and its BR is rather large with
respect to the other non-SM signatures, as we can see in Fig.
\ref{3-4lept} (bottom-left).

\subsection{Signal-to-background analysis}\label{sect:num_analys}

{In this section we perform a signal-to-background analysis
to check the observability at the LHC of some of the signatures discussed 
that may originate from the present $B-L$ model.

In our model setup, wherein the scalar sector is entirely decoupled, 
all interesting $B-L$ signals come from $Z'$ production, 
since the $Z'$ is the only new particle whose couplings to the SM partons
are large. The most efficient hadro-production process involving a
$Z'$ boson is the Drell-Yan (DY) mode
\begin{equation}\label{Z_B-L_prod}
q\bar{q} \rightarrow Z'\, ,
\end{equation}
where $q$ is either a valence-quark or a sea-quark {in the proton}.
At the parton level, the $Z'$ production cross section for
process (\ref{Z_B-L_prod})
depends on two main parameters: $M_{Z'}$ and
$g'_1$. In Fig. \ref{Zpxs} we present the  $Z'$ hadro-production cross section
$\sigma$ at the
LHC as a function of both $M_{Z'}$ and $g'_1$, in the
ranges $0.5$ TeV $< M_{Z'}< 5$ TeV and $0.1 < g'_1 < 0.5$,
respectively, while 
Fig. \ref{zpxs_cont} presents} the contour levels in the ($M_{Z'}$, $g'_1$)
plane for $\sigma = 4$ pb, $0.3$ pb, $50$ fb and $5$
fb. {The shaded area in Fig. \ref{zpxs_cont}
is excluded by eq.~(\ref{LEP_bound}).}
\begin{figure}[!ht]
  \centering
  \subfigure[]{ 
  \label{Zpxs}
  \includegraphics[width=0.5\textwidth]{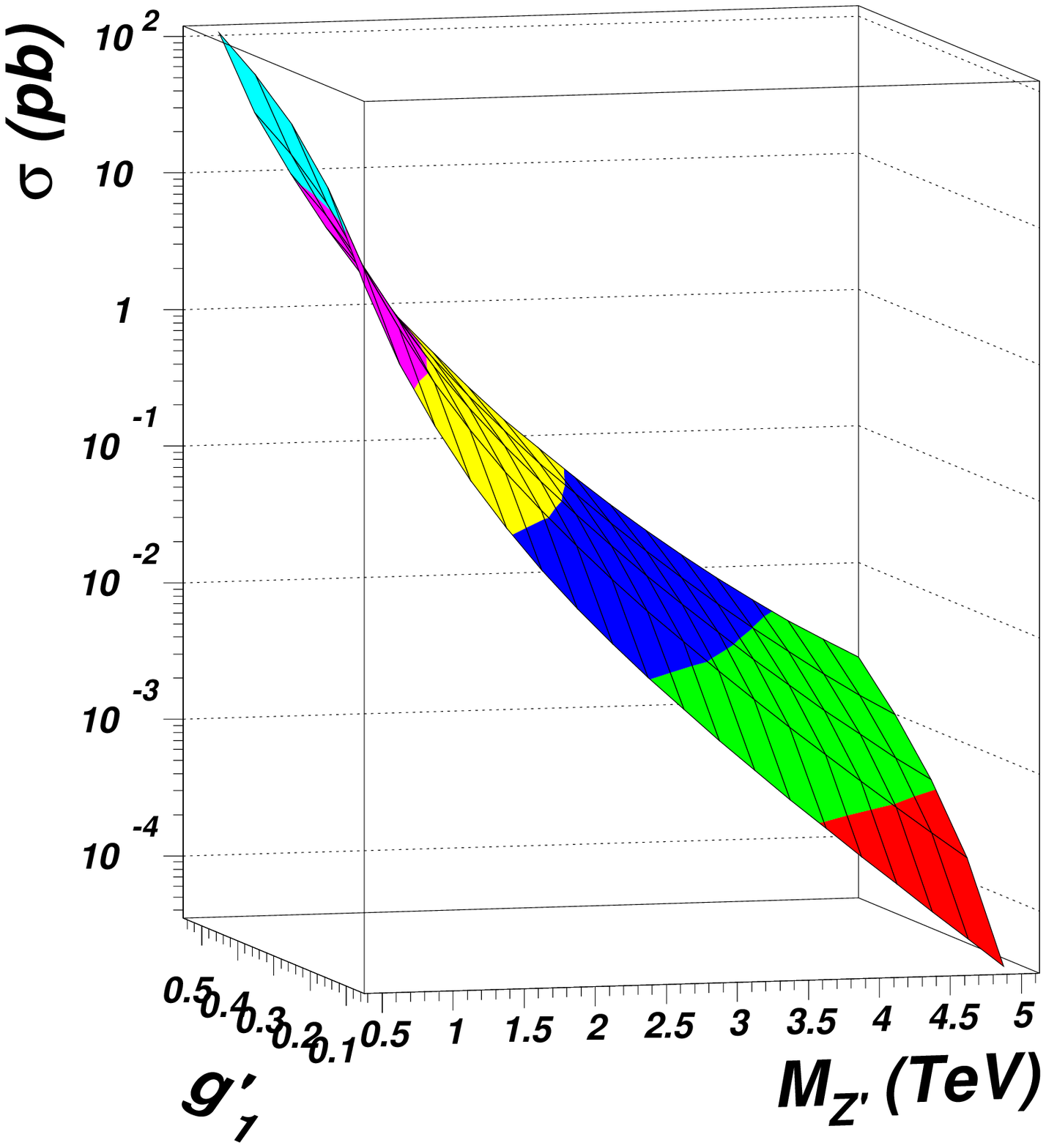}}%
  \subfigure[]{
  \label{zpxs_cont}
  \includegraphics[angle=0,width=0.5\textwidth]{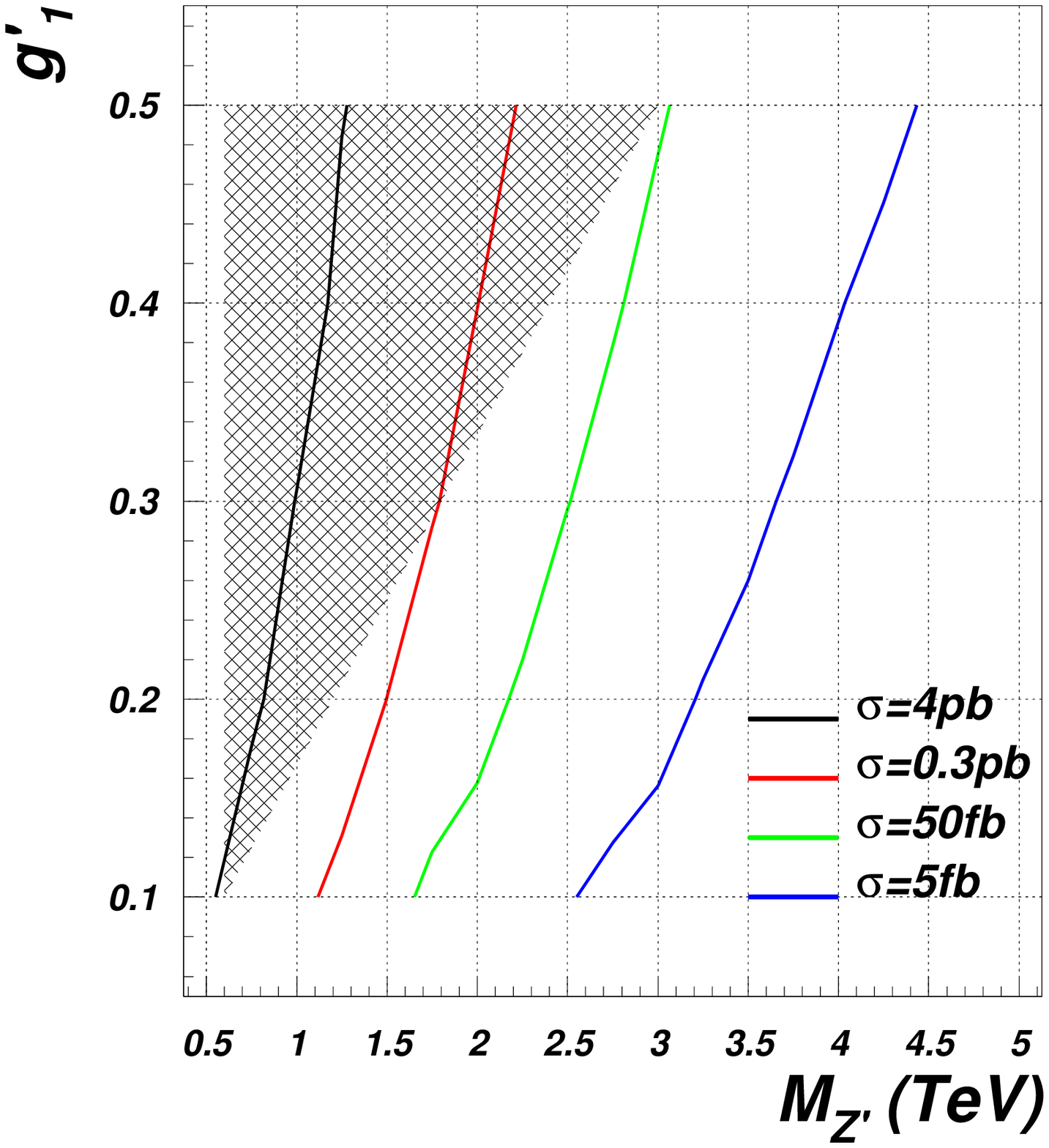}}
  \caption{$Z'$ hadro-production cross section at the LHC over
 the ($M_{Z'}$, $g'_1$) plane:
\ref{Zpxs} as a continuous function of  $M_{Z'}$ and $g'_1$ and
\ref{zpxs_cont} in the form
of contour lines for four fixed values of
production rates. The dark shaded area on the right-hand side plot
is the region excluded by LEP constraints, see eq. (\ref{LEP_bound}).}
  \label{Zp_xs}
\end{figure}

We expect that the LHC will discover a $Z'$ boson from our $B-L$ model in
the standard di-lepton decay channel. 
In Fig. \ref{Zp_prof} we therefore show the $Z'$ line-shape, i.e., the
differential cross section for process  (\ref{Z_B-L_prod}) 
as a function of the invariant mass of its decay
products, e.g., as obtained from the $Z'$ decay into a pair of muons:
\begin{equation}\label{Z_B-L_subproc}
q\bar{q} \rightarrow Z' \rightarrow \mu^+\,\mu^-\, ,
\end{equation}
for the following values of the input parameters: $M_{Z'}=1.5$ TeV,
$g'_1=0.1\div 0.5$ (in 0.1 steps) and $M_{\nu_h}=200$ GeV.
While the di-lepton mode is a powerful $Z'$ discovery channel, its sensitivity to
the presence of heavy neutrinos is however only indirect, through the $Z'$ width, and in fact
very weak, as 
$\Gamma_{Z'}$ varies never more than 20\% or so due to the presence
of the new states (recall Fig.~\ref{Zp_width}(c)).


\begin{figure}[!ht]
\begin{center}
\includegraphics[angle=0,width=0.75\textwidth]{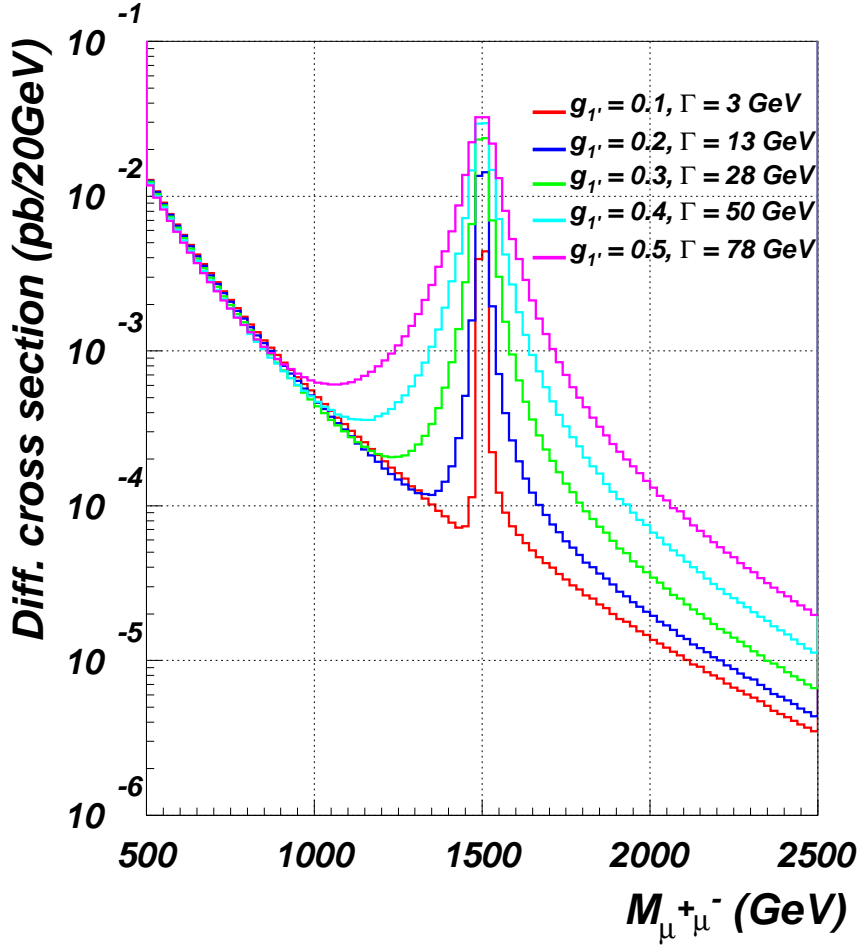}
\caption{Differential cross sections for $q\bar q\to Z'\to\mu^+\mu^-$ 
at the LHC for $M_{Z'}=1500$ GeV, $g'_1=0.1\div 0.5$ (in 0.1 steps)
 and $M_{\nu_h}=200$ GeV as a
function of $M_{\mu ^+\mu^-}$.}
\label{Zp_prof}
\end{center}
\end{figure}



In contrast, multi-lepton signatures carry the hallmark of the heavy neutrinos
as the latter enter directly the corresponding decay chains and these are 
explored here  by performing a detailed 
Monte Carlo (MC) analysis
at the benchmark point $M_Z' = 1.5$ TeV, $g'_1 = 0.2$
and $M_{\nu_h}=200$ GeV. The corresponding total cross section for
$Z'$ production and decay into heavy neutrinos is $46.7$ fb (for
CTEQ6L \cite{CTEQ} with $Q^2=M_{Z'}^2$)\footnote{When discussing event rates in the following,
we will assume an integrated luminosity of $\mathscr{L}=100$
fb${}^{-1}$.}.

Through pairs of heavy neutrinos, other than to fully 
hadronic decays, which are intractable at the LHC
(even in presence of the accompanying missing transverse momentum), the $Z'$ 
can also give rise to $2$-, $3$- or $4$-lepton
signatures, {for}  both $e$ and $\mu$ in the
final state. Amongst the latter, we intend to study here the case of 3-lepton decays. The
reason is twofold. On the one hand, we wish to be able to identify heavy 
neutrino mediation and the presence of only one light neutrino 
in the 3-lepton mode should enable (transverse) mass reconstruction
(contrary to the case of the 4-lepton channel, where
two light neutrinos are involved\footnote{Notice that 
the 4-lepton final state was discussed in \cite{Shabaan}.
See instead Ref.~\cite{Emam:2007dy} for a discussion of
the 2-lepton signature.} On the other hand, we 
ought to minimise
the impact of large backgrounds, so that we neglect here 2-lepton channels
(which could easily by overwhelmed by SM DY and $t\bar t$ production).


When the heavy neutrino decays via the $l^\mp W^\pm$ mode, with a
subsequent leptonic decay of the $W^\pm$, the charged pair of leptons
can carry an invariant mass equal to or lower than the heavy neutrino
mass, with the maximum invariant mass configuration occurring when the
light neutrino is produced at rest, so that the edge in this
distribution corresponds to the $\nu_h$ mass. 
A peak in such a distribution corresponding to the SM-like $Z$ boson,
coming from the $\nu Z$ decay mode for the heavy neutrino will also be
present inthis distribution. The di-lepton invarient mass distribution
is given in Fig.  \ref{3inv}. the difference in the two distributions
illustrates the effect of taking tau lepton decays that produce muons
or electrons into account.

\begin{figure}[!ht]
\begin{center}
\includegraphics[angle=0,width=0.66\textwidth]{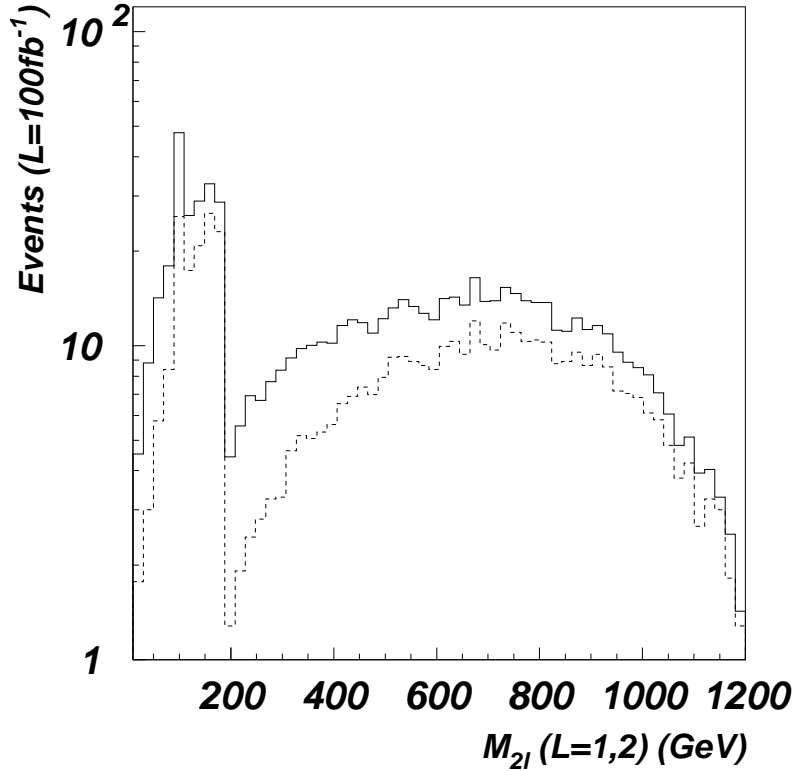}
\caption{Invariant mass of the two most energetic leptons in the $Z' \rightarrow 3l$ decay. The dashed
line refers to data without taking into account the tau lepton. (Here, $\mathscr{L}=100$ fb${}^{-1}$.)}
\label{3inv}
\end{center}
\end{figure}

While the invariant mass distribution can provide some insights into
the mass of the intermediate objects, this is not the best observable
in the case of the $3l$-signature, because the final state neutrino
escapes detection. A more suitable distribution to look at
is the transverse mass defined in \cite{Barger}, i.e.,
\begin{equation}
m^2_T = \left( \sqrt{M^2(vis)+P^2_T(vis)}+\left| \mpt \right| \right) ^2
	- \left( \vec{P^T}(vis) + \mptv\right) ^2\, ,
\end{equation}
where $(vis)$ means {the sum over} the visible particles.  For the
final state considered here we sum over the $3$ leptons and $2$
jets. The transverse mass distribution is shown in figure
\ref{3lep_miss2} where a peak at the $Z'$ mass can be seen. We can
also see evidence for the presence of a heavy neutrino by just
considering the $2$ most energetic leptons and the missing transverse
momentum, since this is the topology relevant to a $\nu_h$ decay.  The
results show that this transverse mass peak for the heavy neutrino is
likely to be the best way to measure its mass. Both of these
configurations are shown in Fig. \ref{3lep_miss2}: the signature of
this model is that both of the above peaks occur simultaneously. (The
different shape for tau-mediated decays is also shown in
Fig. \ref{3lep_miss2} .)

\begin{figure}[!ht]
\begin{center}
\includegraphics[angle=0,width=\textwidth]{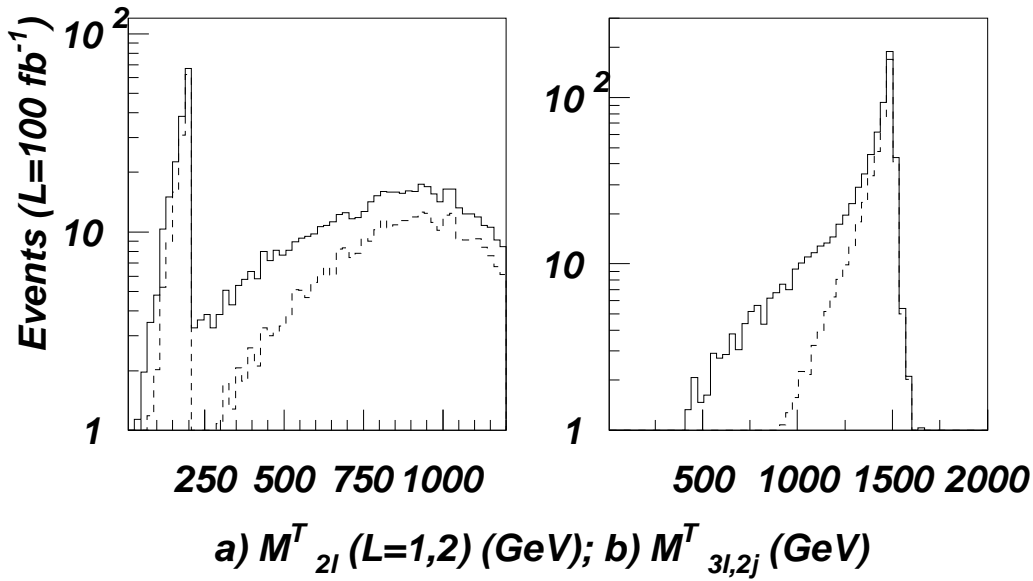}
\caption{The transverse mass of the two most energetic leptons (left) and all the visible particles (right)
in the $Z' \rightarrow 3l$ decay. The dashed line refers to data without taking into account the
tau lepton. (Here, $\mathscr{L}=100$ fb${}^{-1}$.)}
\label{3lep_miss2}
\end{center}
\end{figure}


%
%

The SM background to the 3-lepton signature was
studied using CalcHEP. (For simplicity, from now on, 
we limit ourselves to
the case without leptonically decaying $\tau$'s.)
Making the assumption that the $Z'$ peak has already been identified
and its mass measured elsewhere (as is likely from $Z'\to 2l$
decays)\footnote{Though notice that in Figs. 13--14 the $Z'$ is well
above the background, so that its mass could well be fit -- independently
of process (\ref{Z_B-L_subproc}) --
in the present channel.}, we show that the peak in the $M^T_{2l}$
distribution can be seen despite the initially large backgrounds. 
This enables one to extract a value for $M_{\nu_h}$.
In the evaluation of the background we considered {three} sources (including generation cuts, to improve efficiency):
\begin{itemize}
\item[-] $WZjj$ associated production ($\sigma _{3l} = 246.7$ fb, $l=e,\mu$; $\Delta R_{jj}>0.5$, 
$P^T_{j_{1,2}}>40\mbox{ GeV}$, $\left| \eta_{j_{1,2}} \right| <3)$,
\item[-] $t\overline{t}$ pair production ($\sigma _{2l} = 29.6$ pb, $l=e,\mu$ ($b$-quark not decayed); 
QCD scale $=M_t/2$ to emulate the next-to-leading order cross section; no cuts applied),
\item[-] $t\overline{t}l\nu$ associated production ($\sigma _{3l} = 8.6$ fb, $l=e,\mu$; QCD scale $=\sqrt{\hat s}$, $P^T_l>20$ GeV).
\end{itemize}

In the case of $WZjj$ associated production, three leptons come
from the subsequent leptonic decays of the two gauge
bosons. This is the main source of background.
From $t\overline{t}$ pair production two isolated
leptons come from the decay of the $W^\pm$ produced 
from top decay and
one additional, third lepton,
could come from semileptonic B-meson decay. 
This lepton
though will be not generically  isolated, 
because of the large boost of the $b$-quark
from top-quark decay.
We use this fact  to suppress $t\bar{t}$ background. 
%
%
Finally, $t\overline{t}l\nu$ will produce three isolated leptons
resulting in a significant background despite the small production
cross section.


The first set of cuts we use 
 is designed to impose generic detector angular
acceptances, lepton and jet transverse momentum minimal thresholds and
to provide isolation for leptons and jets: 



\begin{eqnarray}  \nonumber
&&\underbar{\mbox{\large\bf Selection $\#$1}}\\ \nonumber
\left| \eta_{l_{1,2,3}} \right| &<& 2.5,\\ \nonumber
\left| \eta_{j_{1,2}} \right| &<& 3;\\ \nonumber
P^T_{l_1} &>& 15 ~{\rm GeV},\\ \nonumber
P^T_{l_{2,3}} &>& 10 ~{\rm GeV},\\ \nonumber
P^T_{j_{1,2}} &>& 40~{\rm  GeV};\\ \nonumber
\Delta R _{lj}  &>& 0.5\qquad \forall l=1\dots 3, ~j=1,2,\\ \nonumber
\Delta R_{l,l'} &>& 0.2\qquad \forall l,l'=1\dots 3,\\
\Delta R_{j,j} &>& 0.5;
\label{eq:cuts1}
\end{eqnarray}
where
\begin{equation}
\Delta R \equiv \sqrt{\Delta \eta ^2 + \Delta \phi ^2}\nonumber.
\end{equation}


We evaluate the background at two benchmark points for the
signal. For the signal the common parameters are:
\begin{eqnarray} \nonumber
M_{Z'} &=& 1500\mbox{ GeV},\\ \nonumber
g'_1 &=& 0.2 ,\\ 
m_{\nu _l} &=& 10^{-2}\mbox{ eV}
\end{eqnarray}
and two heavy neutrino masses are considered:
%
\begin{eqnarray} \label{Bench_1}
M_{\nu_h} &=& 200\mbox{ GeV},\\ \label{Bench_2}
M_{\nu_h} &=& 500\mbox{ GeV}.
\end{eqnarray}

These two benchmark points provide two kinematically very different examples. 
%
 In the first case the heavy neutrinos are much lighter than the $Z'$
producing highly boosted events. In the second,
their mass is comparable to $M_{Z'}/2$, hence close to their
production threshold, resulting in minimal boost. 
From a merely kinematic point of view, all
other cases will be somewhere between these two.

Special care should be devoted to the treatment of the $t\bar t$ background,
given its large production rates which, however, 
as previously mentioned, can be eliminated
by enforcing a suitable lepton-ject separation.  
The impact of the first set of cuts on the signals and $t\overline{t}$ background
is illustrated in Tab.~\ref{tab-eff}. The $\Delta R _{lj}$ requirement is indeed
extremely effective and reduces this background by a factor of $2 \cdot 10^{-3}$.
Contrary, the loss of signal due to this cut is reasonably small. Also note that the signal
events with the smaller boost have a higher efficiency for passing the 
angular isolation cuts.


\begin{table}[h]
\begin{displaymath}
\begin{array}{|c|cr|cr|cr|} \hline
&&&&&&\\
{\rm Cut} & ~~~~~~~~~~~M_{\nu_{h}}=200 ~{\rm GeV} & 
          & ~~~~~~~~~~~M_{\nu_{h}}=500 ~{\rm GeV} && ~~~~~~~~~ t\overline{t} & \\ 
&
$\#$ {\rm ~of~events} &   {\rm Eff.}~ \%&
$\#$ {\rm ~of~events} &   {\rm Eff.}~ \%&
$\#$ {\rm ~of~events} &   {\rm Eff.}~ \%\\
&&&&&&\\\hline
\mbox{no cuts}		& 482.32(10) 	& 100	& 239.30(10) & 100	&1.28 \cdot 10^{6} 	& 100	\\
\mbox{$\eta$ cuts}	& 346.44(7)	& 71.8	& 170.79(7)  & 71.4	&5.1 \cdot 10^{5} 	& 43.9	\\
\mbox{$\Delta R$+$P_T$ cuts}	& 68.043(15) 	& 14.1	& 73.668(31) & 30.8	&99.699(3)	 	& 0.014	\\ \hline  
\end{array}
\end{displaymath}
\caption{Efficiencies of the Selection $\#$1 cuts
for the two benchmark signals and the 
$t\overline{t}$ background, 
for events with three or more leptons and with two or more jets in the final state
for $\mathscr{L}=100$ fb${}^{-1}$.
In case of $\Delta R_{jj}<0.5$ partons were merged into one `jet'
at the very beginning of the selection.}
\label{tab-eff}
\end{table}
%
%


Figs.~\ref{fig:set1a}--\ref{fig:set1b} show the distributions in
$M_{jj}$, $M^T_{3\ell jj}$, $M_{\ell jj}$ and $M^T_{2\ell}$ after  Selection $\#$1 cuts,
for the signal with the two heavy neutrino
masses, 200 and 500 GeV, that we are considering and the and backgrounds. 

%
\begin{figure}[!ht]
  \includegraphics[angle=0,width=\textwidth]{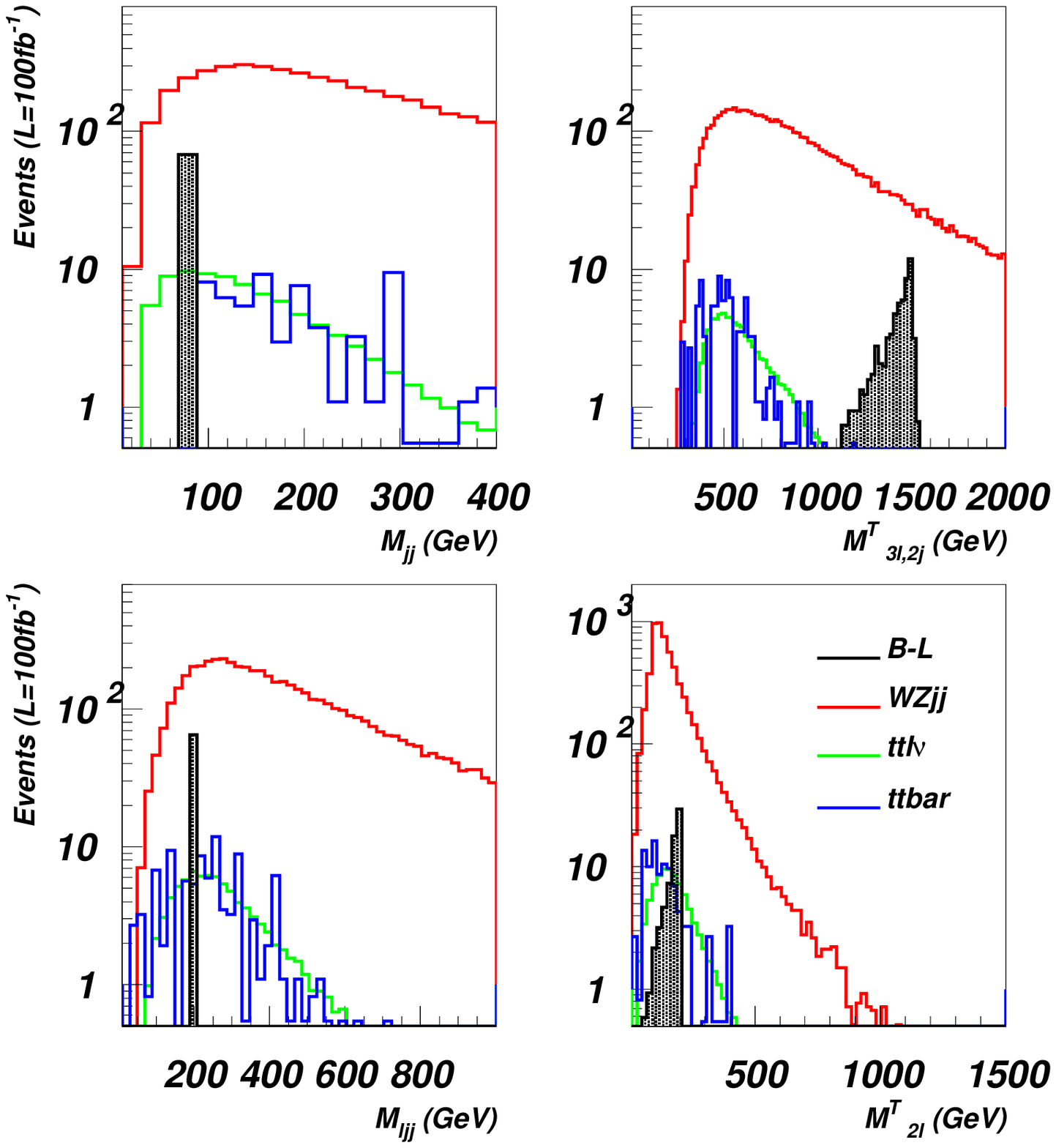}
  \caption{Signal ($M_{\nu_h}=200$ GeV) and background distributions after the 
  Selection $\#$1 cuts.  (Here, $\mathscr{L}=100$ fb${}^{-1}$.)}\label{fig:set1a}
\end{figure}
\begin{figure}[!ht]
  \includegraphics[angle=0,width=\textwidth]{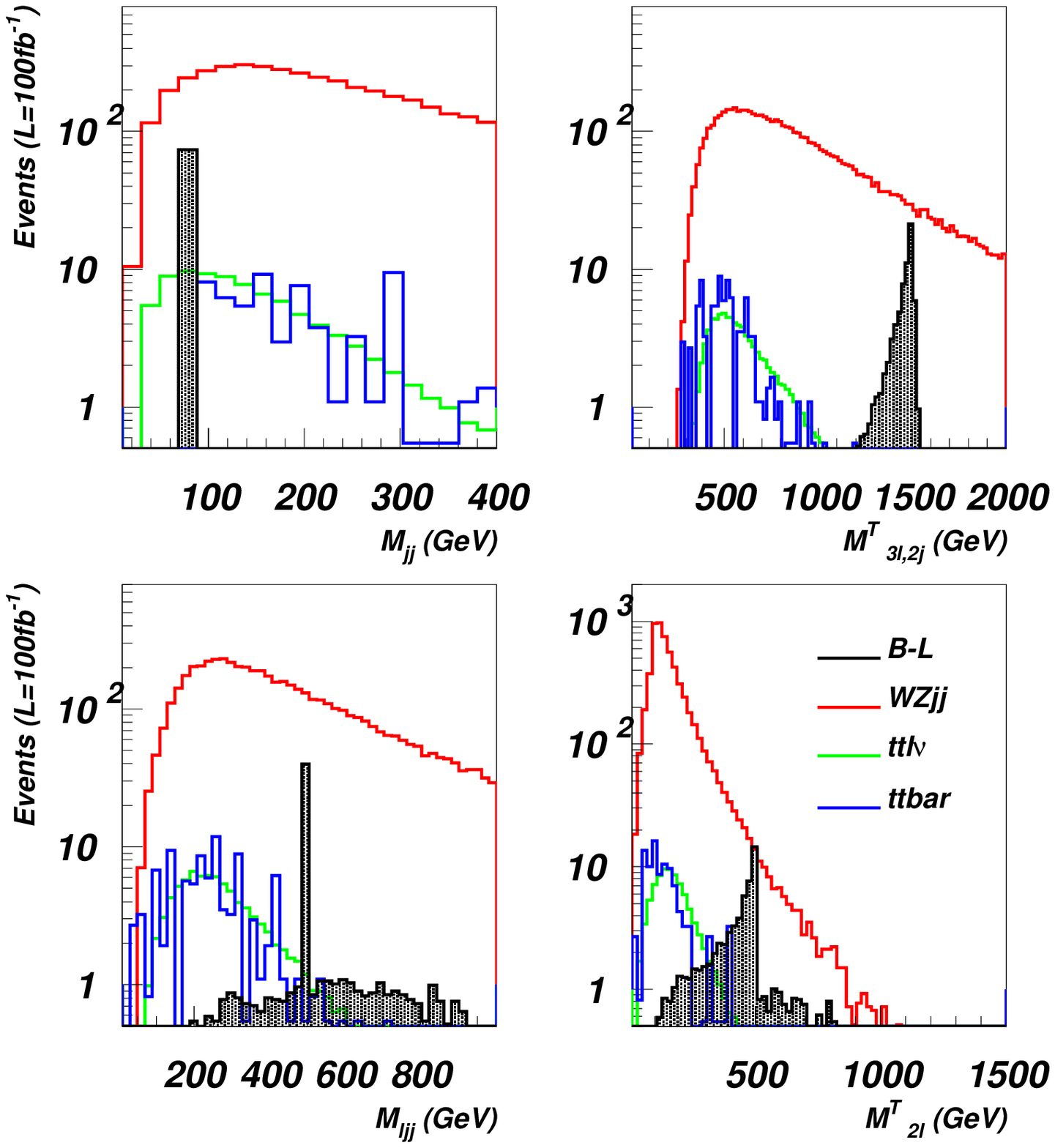}
  \caption{Signal ($M_{\nu_h}=500$ GeV) and background distributions after the  
  Selection $\#$1 cuts.  (Here, $\mathscr{L}=100$ fb${}^{-1}$.)}\label{fig:set1b}
\end{figure}

In signal events both jets come from the $W^\pm$ therefore we apply the
following constraint:
\vspace*{0.15cm}

\centerline{\underbar{\large\bf  Selection $\#$2}}
\begin{equation} \label{cut_2}
\left| M_{jj} - M_W \right| < 20~{\rm GeV}.
\end{equation}
{After the application of this cut the other distributions considered
are shown in Figs.~\ref{fig:set2a}--\ref{fig:set2b} for the 200
and 500 GeV heavy neutrino masses, respectively (here, we now also
show the difference between the $M^T_{2l}$ and $M_{ljj}$
distributions).} From these plots it is clear that transverse mass $M^T_{3\ell
jj}$ provides good discrimination between signal and background. 
The following cut is then used to further suppress the background. 
\vspace*{0.5cm}

\centerline{\underbar{\large\bf  Selection $\#$3}}
\begin{equation} \label{cut_3}
\left| M^T_{3l2j} - M_{Z'} \right| < 250~{\rm GeV}.
\end{equation}

\begin{figure}[!ht]
  \includegraphics[angle=0,width=\textwidth]{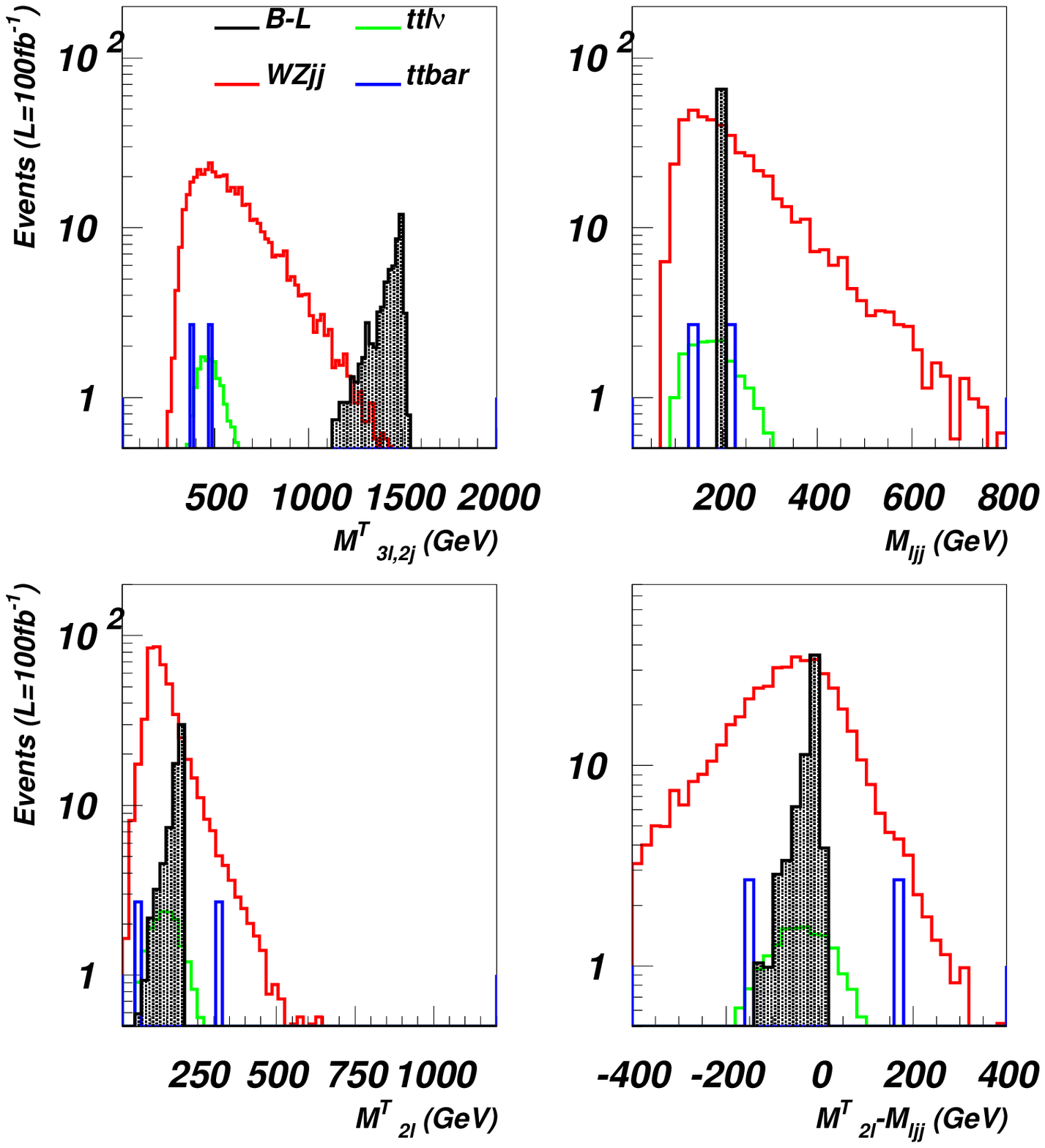}
  \vspace*{-0.5cm}
  \caption{Signal ($M_{\nu_h}=200$ GeV) and background distributions after the 
  Selection $\#$1 and $\#2$ cuts. (Here, $\mathscr{L}=100$ fb${}^{-1}$.)}\label{fig:set2a}
\end{figure}
\begin{figure}[!ht]
  \includegraphics[angle=0,width=\textwidth]{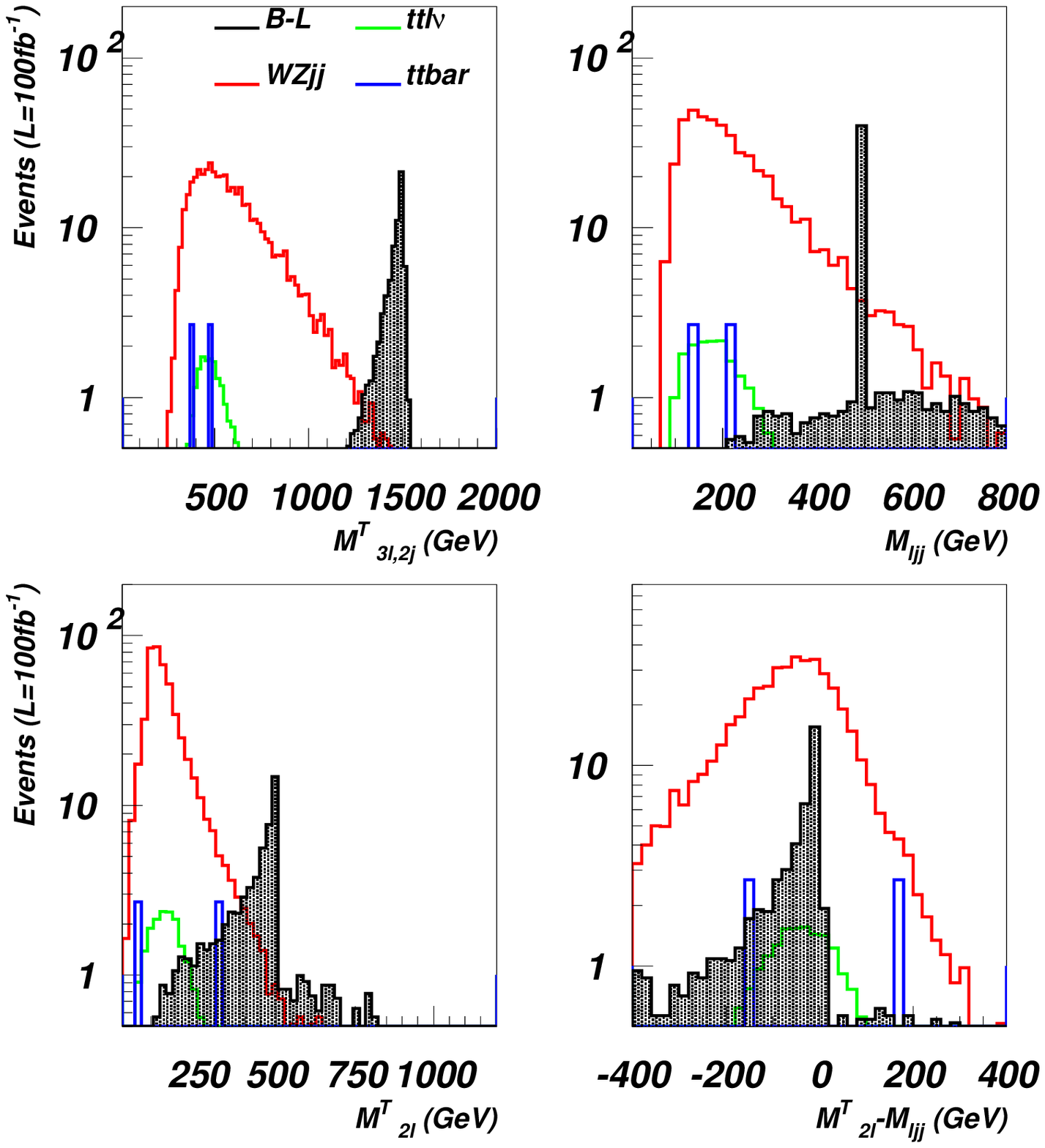}
  \vspace*{-0.5cm}
  \caption{Signal ($M_{\nu_h}=500$ GeV) and background distributions after 
  Selection $\#$1 and $\#2$ cuts. (Here, $\mathscr{L}=100$ fb${}^{-1}$.)}\label{fig:set2b}
\end{figure}

\begin{figure}[!ht]
  \includegraphics[angle=0,width=\textwidth]{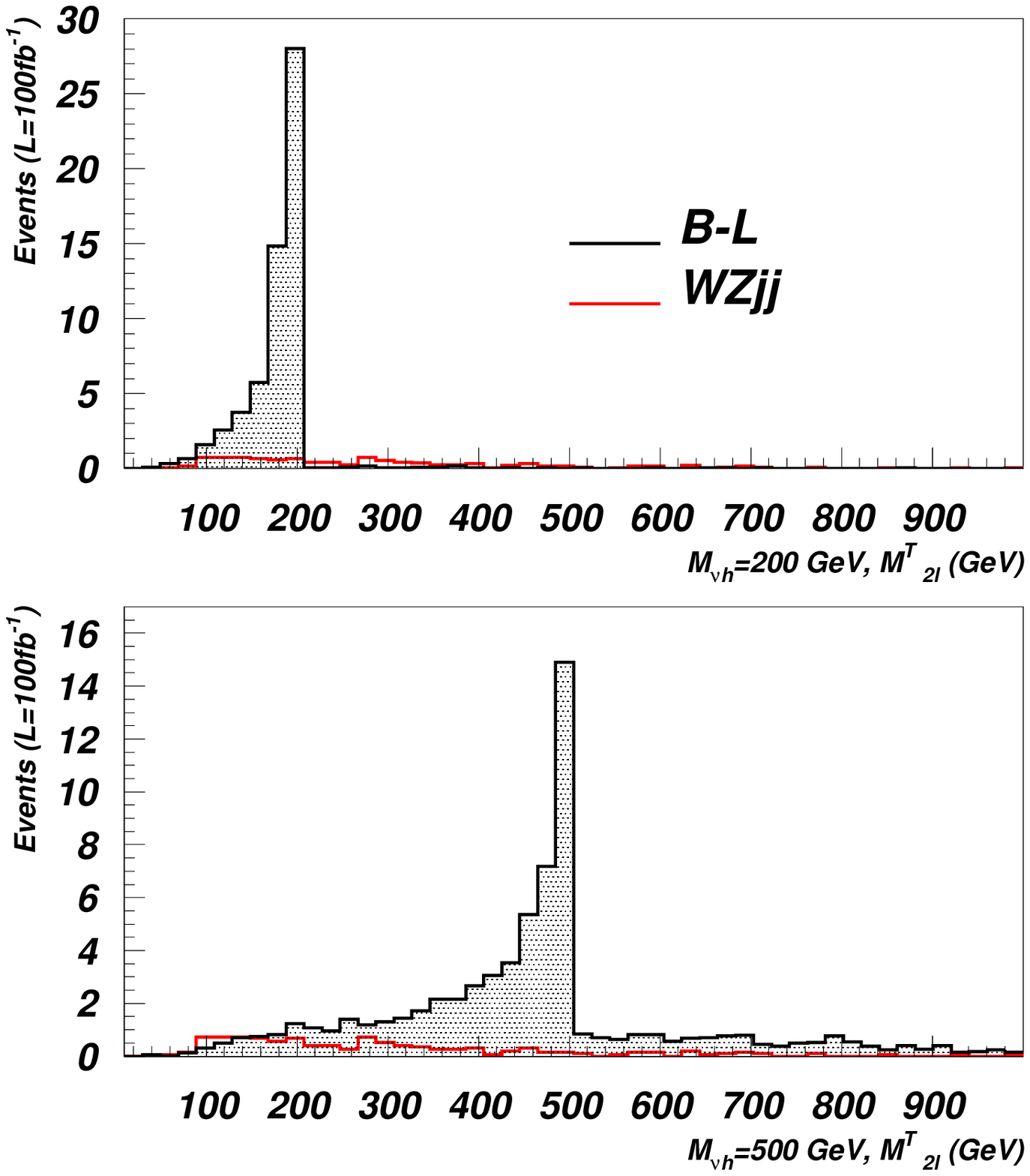}
  \vspace*{-0.5cm}
  \caption{Signal ($M_{\nu_h}=200$ GeV, top, and $M_{\nu_h}=500$ GeV, bottom) 
  and background distributions after the Selection $\#$1, $\#$2 and $\#$3 cuts. 
 (Here, $\mathscr{L}=100$ fb${}^{-1}$.)}\label{fig:set3}
\end{figure}

{After this set of cuts we end up with a very clean signal for both a 200 and
500 GeV $\nu_h$ mass in the di-lepton transverse mass distribution, in
fact practically free from background, as shown in
Fig.~\ref{fig:set3}. Notice that this $M_{2\ell}^T$ variable was
formed from the two closest (in $\Delta R_{ll}$) leptons since they
are likely to originate from the same boosted $\nu_h$ (unlike
Fig.~\ref{3lep_miss2} -- where the two most energetic leptons were
used.)}

{In order to establish the signal we finally select events around the
visible $M^T_{2l}$ peak, by requiring:\\
\vspace*{0.5cm}
\centerline{\underbar{\large\bf  Selection $\#$4}}
\vspace*{-0.45cm}
\begin{equation} \label{cut_4}
0< M^T_{2l} < 250~{\rm{GeV}} \qquad \mbox{ or } \qquad 400~{\rm{GeV}}< M^T_{2l} < 550~{\rm{GeV}},
\end{equation}
depending on the benchmark signal under consideration. The
efficiencies of the  Selection $\#$1--4 cuts, are given in Tab.~\ref{tab:alleff}.
This summary clearly confirms the feasibility of the extraction of
both signals after even less than 100 fb$^{-1}$ of accumulated
luminosity.

\begin{table}
\begin{center}
{$M_{\nu_h}=200$ GeV}
\begin{displaymath}
\begin{array}{|c|cr|cr|cr|cr|c|} \hline
{\rm Cuts} & {\rm Ev.}~{\rm Signal} & {\rm Eff.}~ \% & {\rm Ev.}~ WZjj & {\rm Eff.}~ \% & {\rm Ev.}~t\overline{t} & {\rm Eff.}~ \% & {\rm Ev.}~t\overline{t}l\nu & {\rm Eff.}~ \% & S/\sqrt{B} \\ \hline
1	& 68.043(15) & 100	& 5875.02(24)	 & 100	& 99.699(3)	& 100		& 89.14(16)	& 100	& 0.87 \\
2 	& 68.043(15) & 100 	& 498.83(2) 	 & 8.5	& 5.3822(2)	& 5.4		& 19.38(3)	& 21.8	& 2.97 \\
3	& 58.842(13) & 86.5 	& 10.5755(4) 	 & 12.7	& 0		& 0.8		& 0.0667(1)	& 2.2	& 18.0 \\
4	& 56.038(12) & 94.1	& 4.4881(2)	 & 67.6	& 0		& 56.4		& 0.03047(5)	& 64.8	& 26.3 \\   \hline
\end{array}
\end{displaymath}
\end{center}
\vspace*{0.250cm}
\begin{center}
{$M_{\nu_h}=500$ GeV}
\begin{displaymath}
\begin{array}{|c|cr|cr|cr|cr|c|} \hline
{\rm Cuts} & {\rm Ev.}~{\rm Signal} & {\rm Eff.}~ \% & {\rm Ev.}~ WZjj & {\rm Eff.}~ \% & {\rm Ev.}~t\overline{t} & {\rm Eff.}~ \% & {\rm Ev.}~t\overline{t}l\nu & {\rm Eff.}~ \% & S/\sqrt{B} \\ \hline
1	& 73.668(31)	 & 100	& 5875.02(24)	 & 100	& 99.699(3)	& 100		& 89.14(16)	& 100	& 0.95 \\
2 	& 73.668(31)	 & 100 	& 498.83(2) 	 & 8.5	& 5.3822(2)	& 5.4		& 19.38(3)	& 21.8	& 3.22 \\
3	& 68.833(29) 	 & 93.4 & 10.5755(4) 	 & 12.7	& 0		& 0.8		& 0.0667(1)	& 2.2	& 21.1 \\
4	& 46.337(20)	 & 66.0	& 2.87857(1) 	 & 7.1	& 0		& 8.7		& 0.00952(2)	& 10.1	& 27.6 \\   \hline
\end{array}
\end{displaymath}
\end{center}
\caption{Signal ($M_{\nu_h}=200$ GeV at the top and $M_{\nu_h}=500$ GeV at the bottom)  and
background events per ${\mathscr{L}}=100\;{\rm fb}^{-1}$ and efficiencies following the
sequential application of  Selection $\#$1--4 cuts.}\label{tab:alleff}
\end{table}

\section{Conclusions}\label{sect:conclusions}
We have analyzed the LHC discovery potential in the $Z'$ and heavy
neutrino sector of a (broken) $U(1)_{B-L}$ enlarged SM also
encompassing three heavy (Majorana) neutrinos and found that novel
signals can be established. The most interesting {new signature
involves} three leptons (electron and/or muons), two jets plus missing
transverse momentum {coming from a $Z'$ decay chain into heavy
neutrinos}.  Various mass distributions (both invariant and
transverse) can be used to not only extract the signal after a few
years of LHC running, but also to measure the $Z'$ and heavy neutrino
masses involved. This is possible through DY production and decay via
$q\bar q\to Z'\to \nu_h \nu_h$.  In fact, for a large portion of the
parameter space {of our $B-L$ model}, the heavy neutrinos are rather
long-lived particles, so that they produce displaced vertices in the
LHC detectors, that can be distinguished from those induced by $b$-quarks. 
%
%
%
%
%
In addition, from the simultaneous measurement of both the heavy
neutrino mass and decay length one can estimate the {absolute} mass
of the parent light neutrino, for which at present, only limits
exist. 

This work has used a MC simulation based on a CalcHEP implementation
of the $B-L$ model.  {The analysis has been done at the parton level
though we have verified that our results are stable against the
implementation of typical ATLAS/CMS hadronic and electromagnetic
calorimeter energy 
resolution effects. As benchmark
scenarios of the $B-L$ model we have chosen two that should be
accessible at the LHC, having a $Z'$ mass and fermion couplings not
far beyond the ultimate reach of Tevatron and LEP and displaying two extreme
relative conditions between the $Z'$ and heavy neutrinos, that is, one
with the latter produced at rest and the other highly boosted in the
$Z'$ direction.


\newpage
\section*{APPENDIX: FEYNMAN RULES FOR HEAVY NEUTRINO INTERACTIONS}

In this Appendix, we list the Feynman rules involving the heavy neutrino of
the $B-L$ model considered. The intervening quantities are defined in the
main text. 

\scalebox{1.5}{
\begin{picture}(170,79)(0,30)
\unitlength=1.0 pt
\SetScale{1.0}
\SetWidth{0.1}      
\scriptsize    
\Text(20.0,65.0)[r]{$\nu _h$}
\Line(0.0,60.0)(28.0,60.0) 
\Text(50.0,70.0)[l]{$l$}
\Line(28.0,60.0)(49.0,70.0) 
\Text(50.0,50.0)[l]{$W$}
\Photon(28.0,60.0)(49.0,50.0){2.0}{4}
\Text(30,20)[b] {$\displaystyle \frac{\sqrt{2}e}{4\sin{\vartheta _W}}\sin{\alpha _\nu}$}
\end{picture} 
}%
\scalebox{1.5}{
\begin{picture}(170,79)(0,30)
\unitlength=1.0 pt
\SetScale{1.0}
\SetWidth{0.1}      
\scriptsize    
\Text(20.0,65.0)[r]{$\nu _h$}
\Line(0.0,60.0)(28.0,60.0) 
\Text(50.0,70.0)[l]{$\nu _l$}
\Line(28.0,60.0)(49.0,70.0) 
\Text(50.0,50.0)[l]{$Z$}
\Photon(28.0,60.0)(49.0,50.0){2.0}{4}
\Text(30,20)[b] {$\displaystyle -\frac{e}{4\sin{\vartheta _W}\cos{\vartheta _W}}\sin{2\alpha _\nu}$}
\end{picture} 
}
\\\\
\scalebox{1.5}{
\begin{picture}(170,79)(0,30)
\unitlength=1.0 pt
\SetScale{1.0}
\SetWidth{0.1}      
\scriptsize    
\Text(20.0,65.0)[r]{$\nu _h$}
\Line(0.0,60.0)(28.0,60.0) 
\Text(50.0,70.0)[l]{$\nu _l$}
\Line(28.0,60.0)(49.0,70.0) 
\Text(50.0,50.0)[l]{$H_1$}
\DashLine(28.0,60.0)(49.0,50.0){3.0} 
\Text(70,20)[b] {$\displaystyle \frac{1}{2x}\left( -\sqrt{2}x y^\nu c_\alpha \cos{2\alpha _\nu}+M_{\nu _h} \sin{2\alpha _\nu}s_\alpha\right)$}
\end{picture} 
}%
\scalebox{1.5}{
\begin{picture}(170,79)(0,30)
\unitlength=1.0 pt
\SetScale{1.0}
\SetWidth{0.1}      
\scriptsize    
\Text(20.0,65.0)[r]{$\nu _h$}
\Line(0.0,60.0)(28.0,60.0) 
\Text(50.0,70.0)[l]{$\nu _l$}
\Line(28.0,60.0)(49.0,70.0) 
\Text(50.0,50.0)[l]{$H_2$}
\DashLine(28.0,60.0)(49.0,50.0){3.0} 
\Text(70,20)[b] {$\displaystyle \frac{1}{2x}\left( -\sqrt{2}x y^\nu s_\alpha \cos{2\alpha _\nu}-M_{\nu _h} \sin{2\alpha _\nu}c_\alpha\right)$}
\end{picture} 
}
\\\\
\scalebox{1.5}{
\begin{picture}(170,79)(0,30)
\unitlength=1.0 pt
\SetScale{1.0}
\SetWidth{0.1}      
\scriptsize    
\Text(20.0,65.0)[r]{$\nu _h$}
\Line(0.0,60.0)(28.0,60.0) 
\Text(50.0,70.0)[l]{$\nu _l$}
\Line(28.0,60.0)(49.0,70.0) 
\Text(50.0,50.0)[l]{$Z_{B-L}$}
\Photon(28.0,60.0)(49.0,50.0){2.0}{4}
\Text(30,20)[b] {$\displaystyle g'_1\sin{2\alpha _\nu}$}
\end{picture} 
}
\scalebox{1.5}{
\begin{picture}(170,79)(0,30)
\unitlength=1.0 pt
\SetScale{1.0}
\SetWidth{0.1}      
\scriptsize    
\Text(20.0,65.0)[r]{$\nu _h$}
\Line(0.0,60.0)(28.0,60.0) 
\Text(50.0,70.0)[l]{$\nu _h$}
\Line(28.0,60.0)(49.0,70.0) 
\Text(50.0,50.0)[l]{$Z_{B-L}$}
\Photon(28.0,60.0)(49.0,50.0){2.0}{4}
\Text(30,20)[b] {$\displaystyle g'_1\cos{2\alpha _\nu}$}
\end{picture} 
}
\\\\
\begin{displaymath}
\mbox{where}\ \ \ 
y^\nu = \frac{\sqrt{2 m_{\nu _l} M_{\nu _h}}}{v}\, ,\ \ \ 
\sin{2\alpha _\nu} = -2 \frac{y^\nu \frac{v}{\sqrt{2}}}{\sqrt{4 (y^\nu \frac{v}{\sqrt{2}})^2+M_{\nu _h}^2}}\, , \ \ \ 
\cos{2\alpha _\nu} = \frac{M_{\nu _h}}{\sqrt{4(y^\nu \frac{v}{\sqrt{2}})^2+M_{\nu _h}^2}}.
\end{displaymath}

\end{document}